\documentclass[onecolumn,11pt,a4paper,aps]{revtex4}

\newcommand{\dd}{\mathrm{d}}
\newcommand{\ee}{\mathrm{e}}

\usepackage{amsmath}
\usepackage{graphics}
\usepackage{graphicx}
\usepackage{psfrag}

\begin{document}

\title{Free-fermion entanglement and spheroidal functions}
\author{ Viktor Eisler$^1$ and Ingo Peschel$^2$}
\affiliation{
$^1$Fakult\"at f\"ur Physik, Universit\"at Wien, Boltzmanngasse 5,
A-1090 Wien, Austria\\
$^2$Fachbereich Physik, Freie Universit\"at Berlin,
Arnimallee 14, D-14195 Berlin, Germany
}

\begin{abstract}
We consider the entanglement properties of free fermions in one
dimension and review an approach which relates the problem to the
solution of a certain differential equation. The single-particle eigenfunctions
of the entanglement Hamiltonian are then seen to be spheroidal functions
or generalizations of them. The analytical results for the eigenvalue 
spectrum agree with those obtained by other methods. In the continuum
case, there are close connections to random matrix theory.
\end{abstract}
\maketitle

\section{Introduction}

 The entanglement properties of many-particle quantum states have been the topic
of many studies in recent years \cite{Calabrese/Cardy/Doyon09}. This holds in particular 
for the ground state of free fermionic systems. These are Slater determinants, and in this 
case the reduced density matrix (RDM) for some portion of the total system has the form 
\begin{equation}
\rho = \frac{1}{Z}\; e^{-\mathcal{H}}
\label{rhogen}
\end{equation}
where $\mathcal{H}$ is again a free-particle Hamiltonian, see \cite{review09}.
Its single-particle eigenfunctions $\varphi_k$ and the corresponding eigenvalues
$\varepsilon_k$ can be determined either from the one-particle correlation functions
\cite{Peschel03,Vidal03,Latorre04,Cheong/Henley04} or from the overlap of the occupied states 
in the subsystem \cite{Klich06,CMV11a,CMV11b,CMV12}. Their properties for the most studied 
case, namely a segment in a chain of free fermions, are relatively well-known. Thus the 
low-lying $|\varepsilon_k|$ vary roughly linearly with $k$ and the slope is proportional to
$1/\ln L$ if the subsystem has length $L$ and $\ln L$ is large. This leads
to a logarithmic variation $S =1/3 \ln L$ of the entanglement entropy $S$ which is
characteristic for critical systems and follows from conformal invariance, see \cite{CC09}.
The law was first obtained from an asymptotic analysis of the correlation matrix using
the Fisher-Hartwig conjecture \cite{Jin/Korepin04}, and further subleading terms have
been derived in the same way \cite{Calabrese/Essler10,Fagotti/Calabrese11}. The 
eigenfunctions $\varphi_k$
are largest near the boundaries for small $|\varepsilon_k|$ and concentrated in the
interior of the subsystem for large $|\varepsilon_k|$. This can be seen easily from 
numerical calculations. For a half-filled system, there are also analytical expressions 
in the continuum limit \cite{Peschel04}. Altogether, a picture emerges, where the 
(entanglement) Hamiltonian $\mathcal{H}$ has both bulk and boundary-like states with 
the latter dominating the entanglement.

  Given the relatively simple form of the correlations (or the overlap), one wonders 
whether a complete analytical treatment of the problem is possible. This is, in fact,
the case and it turns out that it was done a long time ago by Slepian in the analysis 
of time and band limited signals \cite{Slepian65,Slepian78}. A good account of the 
background can be found in his 1982 John von Neumann lecture \cite{Slepian83}. The
work has received a large number of citations in very different areas. The analogy 
to the entanglement problem, where one deals with limited regions in momentum space 
(the Fermi sea) and in real space (the subsystem), was pointed out already by Gioev 
and Klich \cite{Gioev/Klich06}. However, while they derived a formula for $S$ in higher 
dimensions, we focus here on the eigenvalue problem. 

 It turns out that the eigenfunctions $\varphi_k$, or their Fourier transforms, are the 
(prolate) spheroidal functions which appear if one separates the Helmholtz equation 
in elliptical coordinates, or 
generalizations of them. They enter in the continuum case because the integral kernel 
of the entanglement problem commutes with the corresponding differential operator. 
Those concentrated in the interior of the subsystem are quite familiar objects, namely 
oscillator functions. They become more complicated once they touch the boundary. In any 
case, the machinery of solving differential equations can be used to obtain their form 
and also the eigenvalues $\varepsilon_k$. In particular, an asymptotic analysis is
possible and gives the low-lying ones for the case of a large subsystem. This leads 
to exactly the same density of states as found in the Fisher-Hartwig approach 
\cite{Suesstrunk/Ivanov12,Jin/Korepin04}. 
If either momenta or positions become discrete, the situation is still similar. Working
with the other, continuous quantity, one has a commuting differential operator and can 
discuss its eigenfunctions. 
One should mention that a lot of this material also appears in random
matrix theory, because in the Gaussian unitary ensemble the same integral kernel enters
and the eigenvalues $\varepsilon_k$ determine the functions of interest, see \cite{Mehta04}.
The connection to this field was pointed out by Keating and Mezzadri \cite{Keating/Mezz04,
Keating/Mezz05} and also used in \cite{Calabrese/Essler10}, but again with the focus on
the entropy.

  The purpose of this paper is to draw attention to the results described above, because
they complement the usual approaches and put the entanglement problem into a broader
context. Our own contribution is mainly the collection and the presentation, including
a number of figures. In section 2 we give a brief general outline of the determination 
of $\varphi_k$ via correlation and overlap matrices. In section 3 we discuss the case 
of an infinite continuum system, where the simple sine kernel and the usual spheroidal 
functions appear, of which we show some examples. Section 4 treats the case of a finite 
continuum system, where the correlations still lead to an integral equation. Section 5
deals with an infinite lattice where the same equation appears in momentum space.
In this case we also discuss the matrix which commutes with the correlation matrix, as well 
as the dispersion relation of the $\varepsilon_k$. Finally, section 6 contains remarks 
on semi-infinite systems and higher dimensions and a brief conclusion is given in Section 7.

\section{Correlation matrices and overlap matrices}

  The structure of the problem is best seen from the formulae for a general 
lattice system with a discrete set of orthonormal single-particle functions $\Phi_q(n)$. 
A many-particle state $|F\rangle$ in which a set $F$ of these is occupied (the Fermi sea) then 
is
\begin{equation}
|F\rangle \;  = \, \prod_{q \in F} \, \,c_q^{\dagger} \,|0 \rangle
\label{state}
\end{equation}
where $c_q^{\dagger}$ are creation operators and $|0 \rangle$ is the vacuum.
With
\begin{equation}
c_n = \sum_q \Phi_q(n) c_q
\label{operators}
\end{equation}
the correlation matrix in state $|F\rangle$ becomes
\begin{equation}
C_{mn} = \langle F|c_m^{\dagger}c_n|F\rangle = \sum_{q \in F}  \Phi_q^*(m)\Phi_q(n)  
\label{corr1}
\end{equation}
and the RDM in a subsystem $S$ is determined by the eigenvalue problem
\begin{equation}
\sum_{j \in S}  C_{ij} \,\varphi_k(j) = \zeta_k \,\varphi_k(i) , \quad i \in S  
\label{corr2}
\end{equation}
If $S$ consists of $L$ sites, (\ref{corr2}) gives $L$ single-particle eigenfunctions 
$\varphi_k$. The eigenvalues $\zeta_k$ with $0 \le \zeta_k \le 1$ are their non-integer 
occupation numbers and related to the $\varepsilon_k$ via
\begin{equation}
\varepsilon_k = \ln \frac {1-\zeta_k}{\zeta_k} \quad \, \mathrm{or} \quad \, 
\zeta_k = \frac{1}{e^{\varepsilon_k}+1} 
\label{epsilon}
\end{equation}
These quantities can also be obtained by constructing the Schmidt decomposition of $|F\rangle$ 
directly \cite{Klich06}. In this case, one forms new orthonormal functions $\Psi_k$ from
the occupied $\Phi_q$ which are orthogonal also in the subsystem (and in the remainder $R$). 
This is done by calculating the overlap matrix
\begin{equation}
A_{qq'} = \sum_{j \in S} \Phi_q^*(j)\Phi_{q'}(j)  
\label{overlap1}
\end{equation}
and solving the eigenvalue problem
\begin{equation}
\sum_{q' \in F} A_{qq'} \, \varphi_k(q') = \zeta_k \, \varphi_k(q) , \quad q \in F  
\label{overlap2}
\end{equation}
Then 
\begin{equation}
\Psi_k(i)= \sum_{q \in F} \varphi_k(q) \Phi_q(i) 
\label{overlap3}
\end{equation}
satisfies (\ref{corr2}) when restricted to $S$. Similarly, its restriction to $R$ 
satisfies the analogue of (\ref{corr2}) with eigenvalue $(1-\zeta_k)$. Moreover, it  
has norm $\zeta_k$ in $S$ and norm $1-\zeta_k$ in $R$. Therefore, if 
$\varphi_k$ is normalized in $S$, one has
\begin{equation}
\varphi_k(i) = \frac{1}{\sqrt{\zeta_k}} \Psi_k(i) , \quad i \in S  
\label{overlap3}
\end{equation}
which gives the pair of relations
\begin{equation}
\varphi_k(i) = \frac{1}{\sqrt{\zeta_k}} \sum_{q \in F} \varphi_k(q) \Phi_q(i)  
  \quad ,  \quad \quad
\varphi_k(q) = \frac{1}{\sqrt{\zeta_k}} \sum_{i \in S} \varphi_k(i) \Phi^{*}_q(i)
\label{mutual}
\end{equation}
The normalization properties allow to obtain the eigenvalues $\zeta_k$ solely from the 
$\Psi_k$ if these can be found by some other means (as will be the case below). If one 
arranges the $\zeta_k$ in decreasing order, the corresponding $\Psi_k$ are less and less
concentrated in $S$. For $L$ sites in the subsystem, (\ref{overlap2}) can have only $L$ 
non-trivial solutions. If the particle number, i.e. the number of states in $F$, exceeds $L$, 
the remaining $\Psi_k$ have no components in the subsystem and $\zeta_k=0$.

In the following we apply these formulae to the case of plane waves, 
$\Phi_q(n) \sim e^{iqn}$ or $\Phi_q(x) \sim e^{iqx}$. The two eigenvalue problems 
(\ref{corr2}) and (\ref{overlap2}) then correspond to working in real space and in 
momentum space, respectively. Moreover, the state $|F\rangle$ will be a normal Fermi sea 
($|q| \le q_F$) and the subsystem $S$ a single segment. The two functions $\varphi_k(i)$ 
and $\varphi_k(q)$ are then related by a Fourier transform limited to a window in $q$ 
space and in real space.

\section{Infinite continuous system}

We now consider free fermions on an infinite line with all momentum states $|q| \le q_F$
filled. The system then has average density $\bar n=q_F/\pi$ and the correlation matrix is
\begin{equation}
C(x-x') = \int_{-q_F}^{q_F} \frac{dq}{2\pi} \, e^{-iq(x-x')} =  
           \frac{\sin q_F(x-x')}{\pi (x-x')} 
\label{corr_cont1}
\end{equation}
Choosing the subsystem $S$ as the segment $-\ell/2 \le x \le \ell/2$, the eigenvalue problem
(\ref{corr2}) becomes 
\begin{equation}
\int_{-\ell/2}^{\ell/2} dx' \, C(x-x') \varphi_k(x') = \zeta_k \,\varphi_k(x) 
\label{corr_cont2}
\end{equation}
or in reduced variables $y=2x/\ell$ with $\varphi_k(\ell y/2)=\psi_k(y)$
\begin{equation}
\int_{-1}^{1} dy' \, K(y-y') \psi_k(y') = \zeta_k \,\psi_k(y) 
\label{corr_cont3}
\end{equation}
with the \emph{sine kernel}
\begin{equation}
K(y-y') = \frac{\sin c(y-y')}{\pi (y-y')}\, , \quad \quad \quad c= q_F \ell/2
\label{sine_kernel}
\end{equation}
Similarly, the overlap matrix is
\begin{equation}
A(q-q') = \int_{-\ell/2}^{\ell/2} \frac{dx}{2\pi} \, e^{-i(q-q')x} =  
           \frac{\sin (q-q')\ell/2}{\pi (q-q')} 
\label{overlap_cont1}
\end{equation}
and with $p=q/q_F$ the eigenvalue problem (\ref{overlap2}) becomes 
\begin{equation}
\int_{-1}^{1} dp' \, K(p-p') \psi_k(p') = \zeta_k \,\psi_k(p) 
\label{overlap_cont2}
\end{equation}
Thus the equations in real space and in momentum space are identical. In the 
following, we will work in real space.

The sine kernel is the ``square'' of another, even simpler symmetric kernel 
$\tilde K(y,y')$
\begin{equation}
K(y-y') = \int_{-1}^{1} dz \,\tilde K(y,z) \tilde K^*(z,y') \, , \quad \quad
\tilde K(y,z) = \sqrt{\frac{c}{2\pi}} \ee^{icyz}
\label{Ktilde}
\end{equation}
Therefore, the eigenvalues $\zeta_k$ of the sine kernel follow from
the eigenvalues $\mu_k$ of $\tilde K$ via $\zeta_k = |\mu_k|^2$.  
Moreover, $\tilde K$ commutes with the second-order \emph{differential operator} \cite{Ince56}
\begin{equation}
D = - \frac{\dd}{\dd y} (1-y^2) \frac{\dd}{\dd y} + c^2y^2 
\label{diff_op}
\end{equation}
provided the functions one operates on are regular at $y =\pm 1$. Therefore the (real)
eigenfunctions of $\tilde K$ and $K$ can be obtained from
\begin{equation}
D \psi = \theta \psi
\label{diff_eigen}
\end{equation}
This equation has continuous and bounded solutions for all $y$ only for a discrete set 
of values $\theta = \theta_k$. The corresponding $\psi_k$ are the prolate spheroidal
wave functions of the first kind and order $m=0$. In the notation of \cite{AS64} these are
called angular functions $S_{0k}$ for $y^2 < 1$ and radial functions $R_{0k}$ for $y^2 > 1$ 
(which fits with our notations of subsystem and remainder). Both are normalized separately 
and connected by joining formulae. Their properties have been investigated in great detail, 
see e.g. \cite{Meixner/Schaefke54,Flammer57,Meixner/Schaefke/Wolf80}.
Studying the solution of (\ref{diff_eigen}) for 
\emph{all} $y$ gives the quantity $\Psi_k$ of (\ref{overlap3}). Of particular interest is 
the case where the subsystem contains many particles. Since the average number is
\begin{equation}
\bar N = \ell \bar n = \frac{2}{\pi} c
\label{part_number}
\end{equation}
this corresponds to large $c$.

\begin{figure}[htb]
\center
\includegraphics[width=0.3\textwidth]{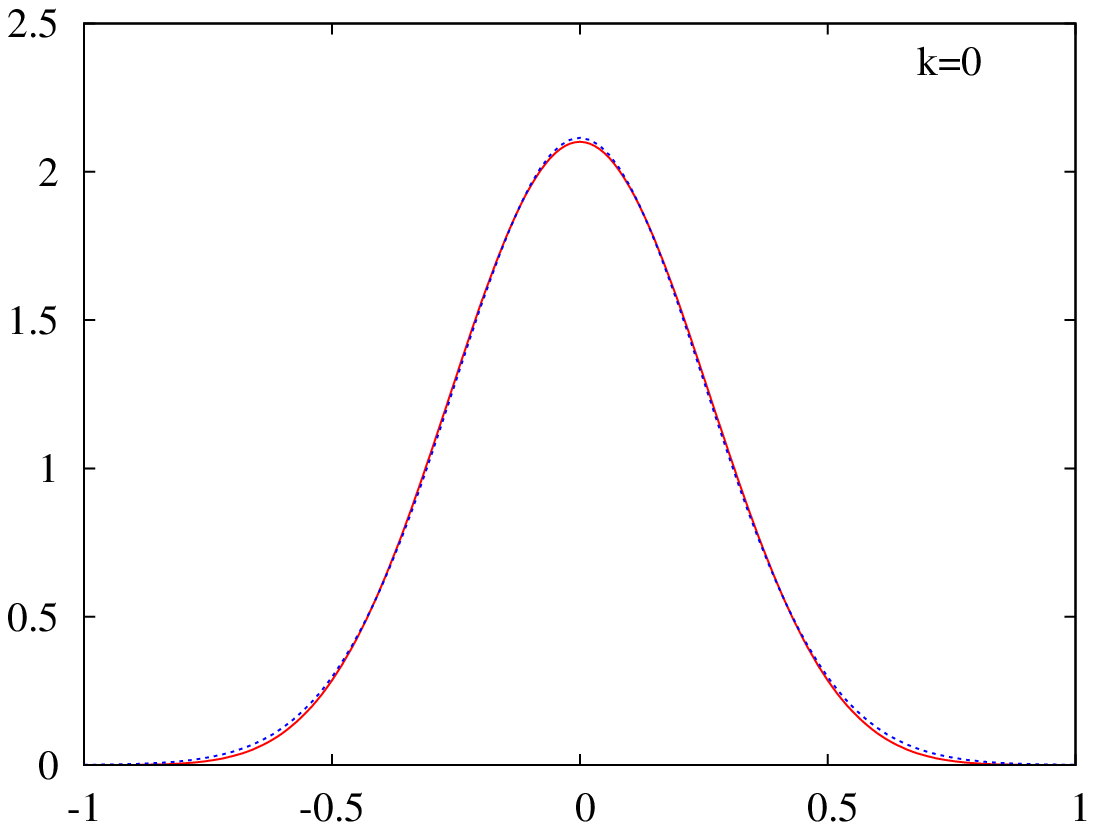}
\includegraphics[width=0.3\textwidth]{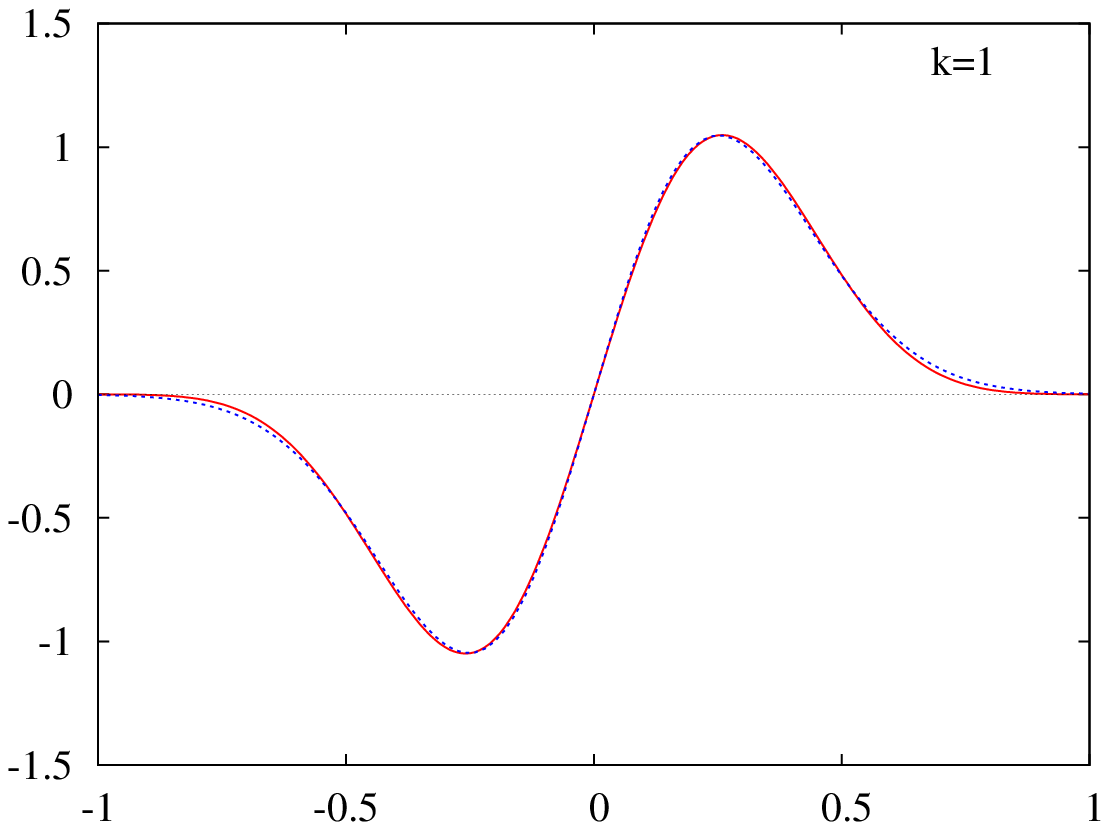}
\includegraphics[width=0.3\textwidth]{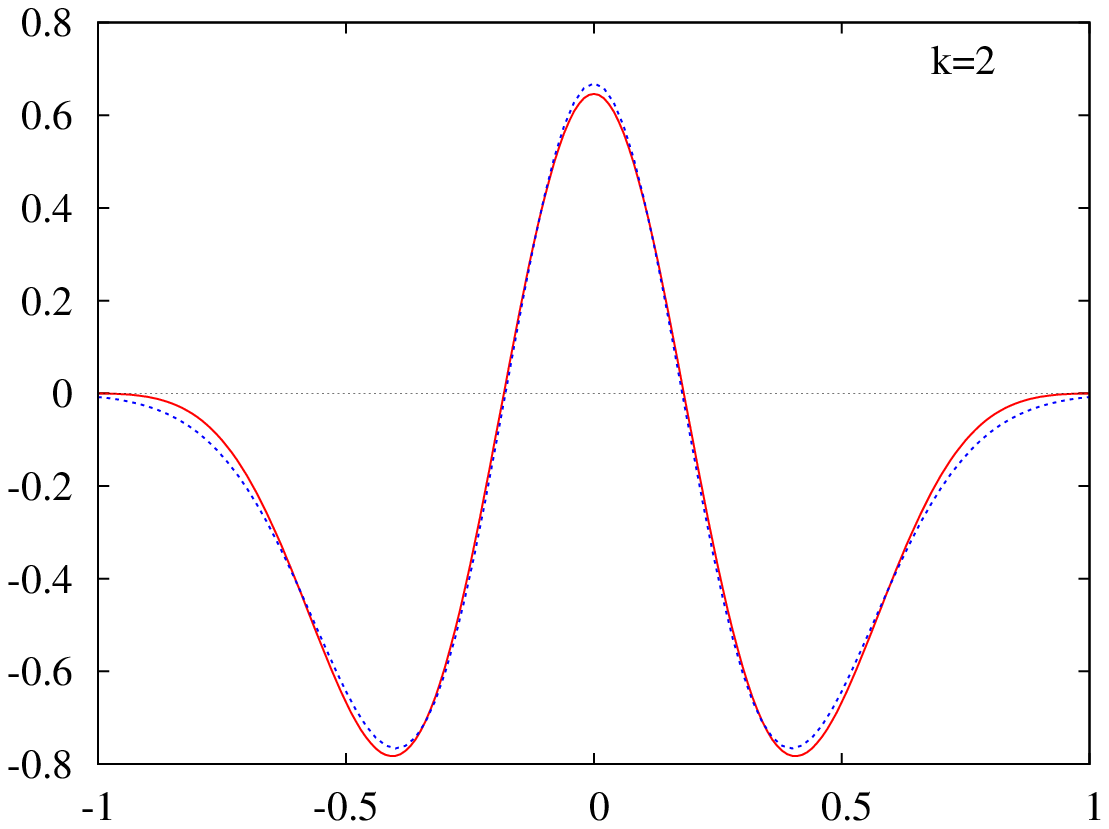}
\includegraphics[width=0.3\textwidth]{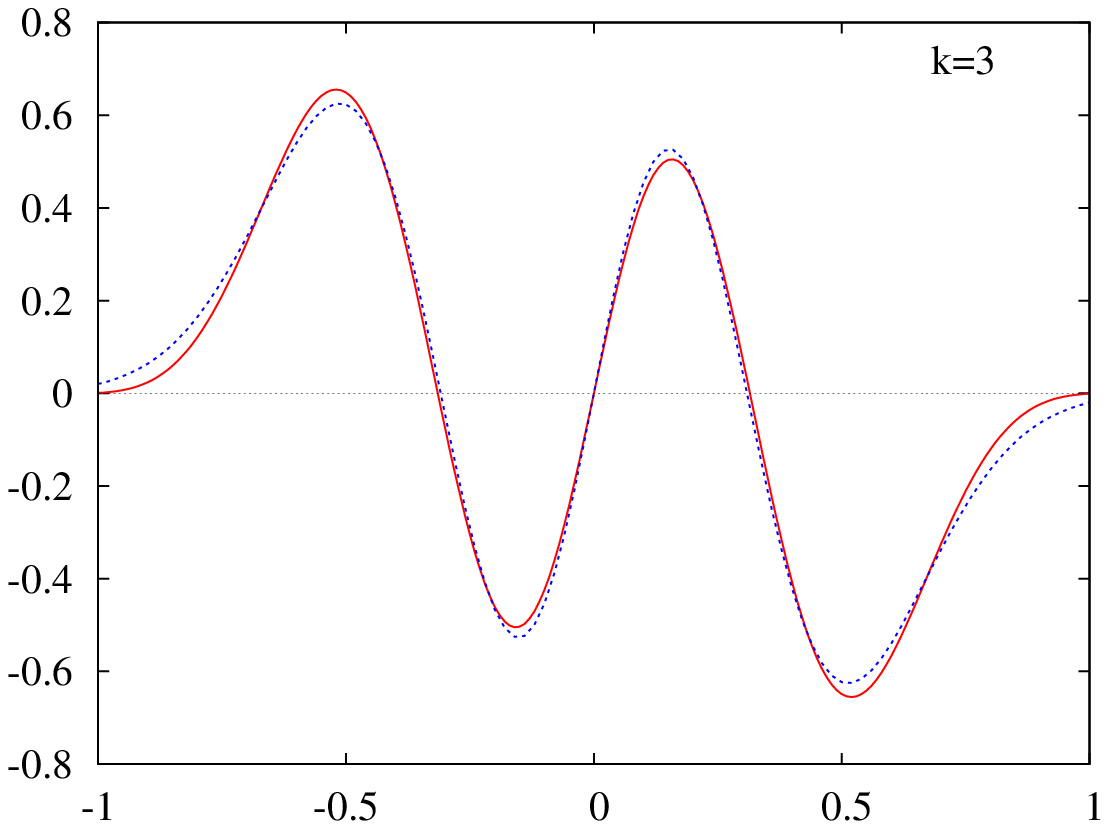}
\includegraphics[width=0.3\textwidth]{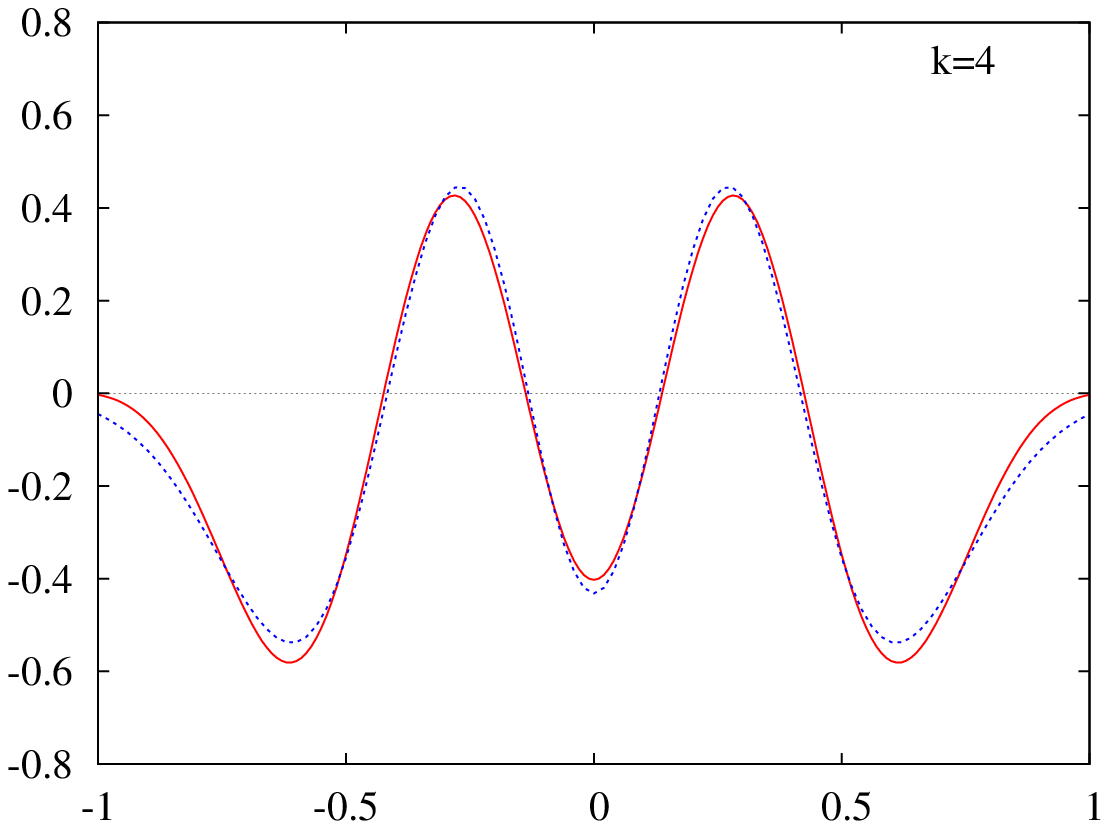}
\includegraphics[width=0.3\textwidth]{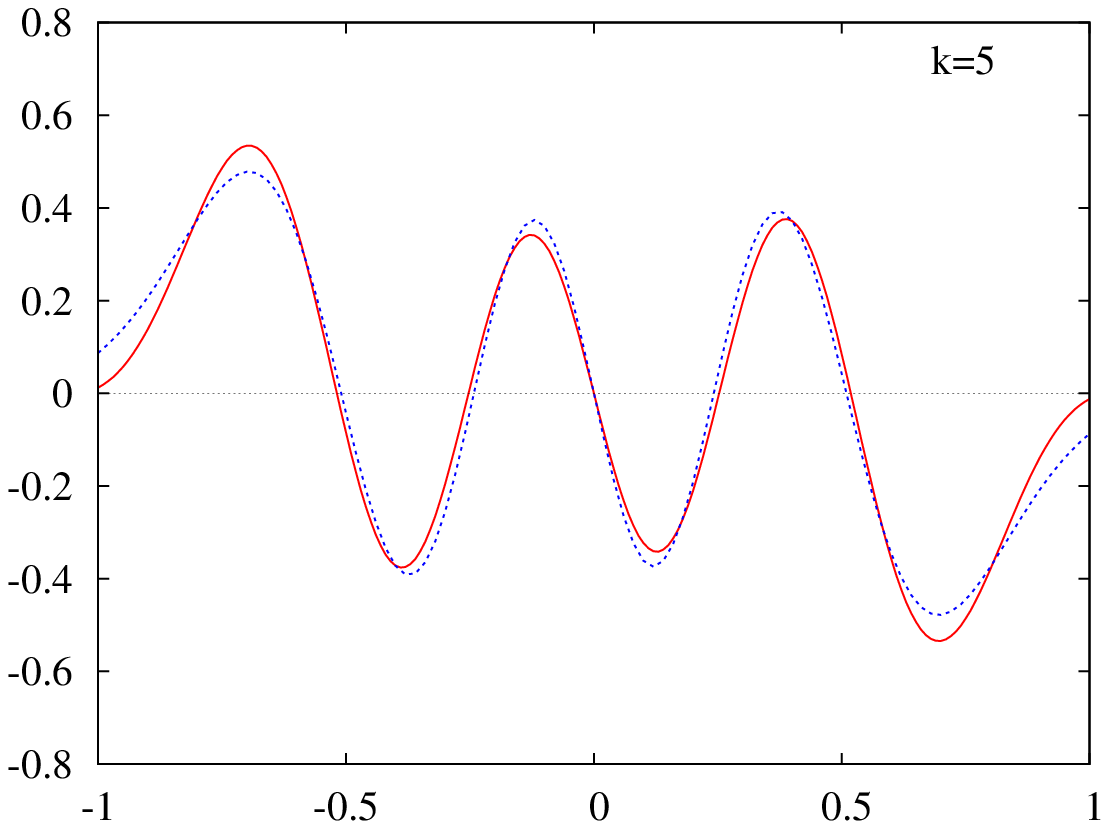}
\includegraphics[width=0.3\textwidth]{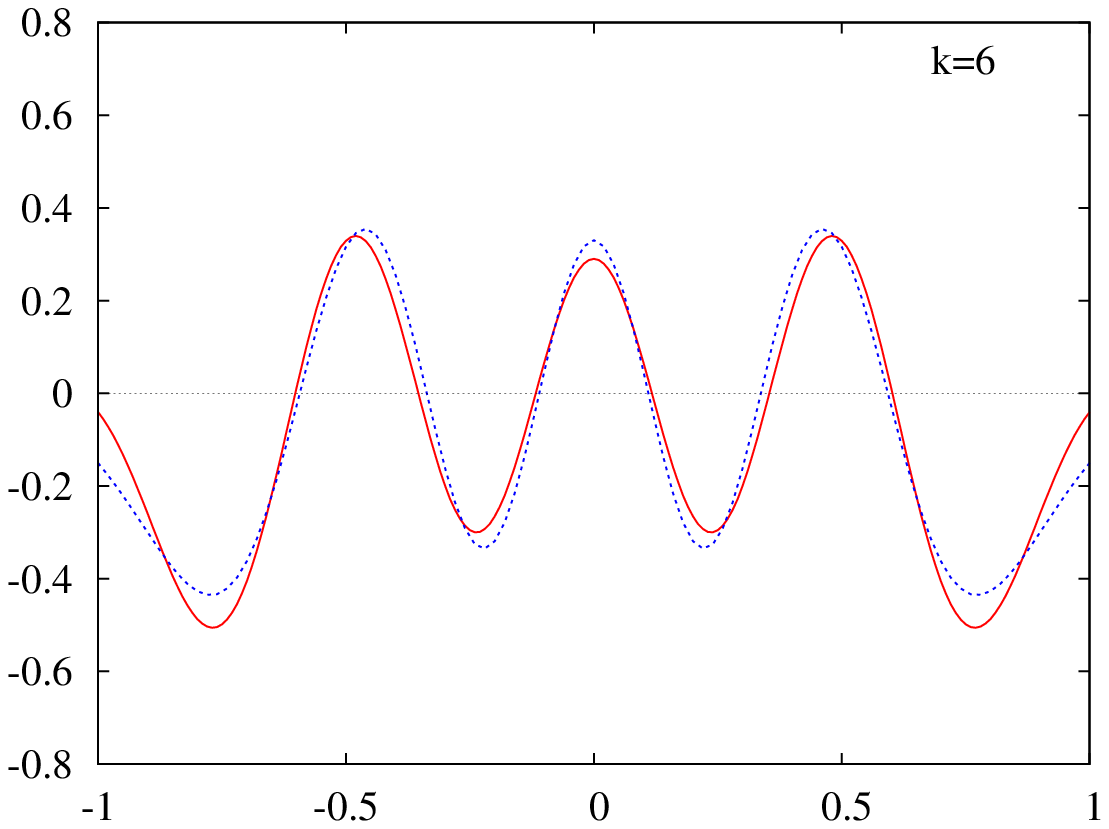}
\includegraphics[width=0.3\textwidth]{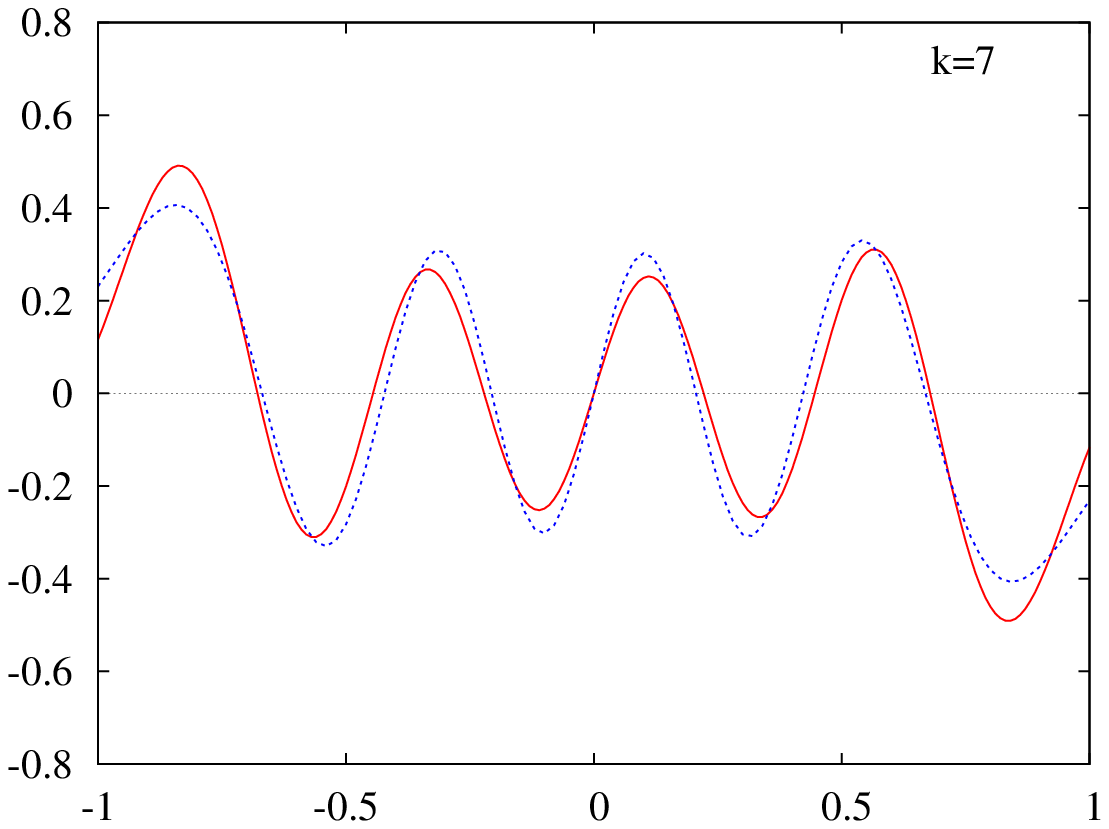}
\includegraphics[width=0.3\textwidth]{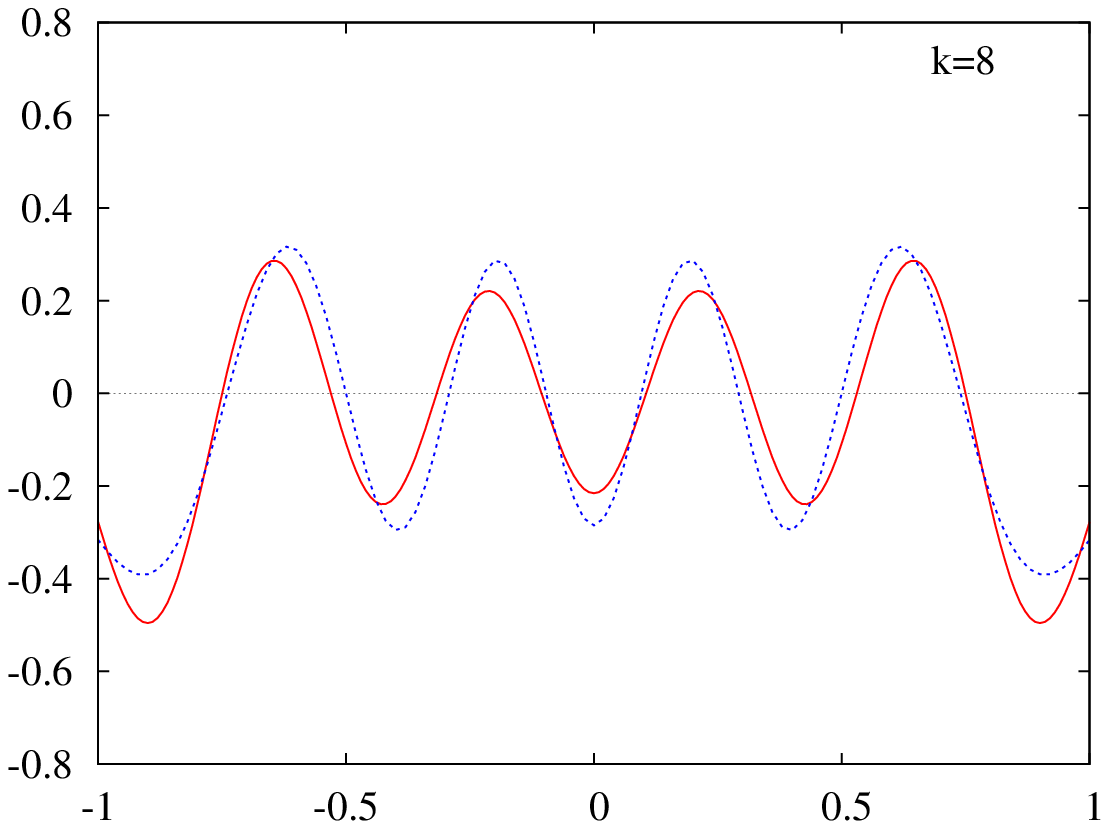}
\includegraphics[width=0.3\textwidth]{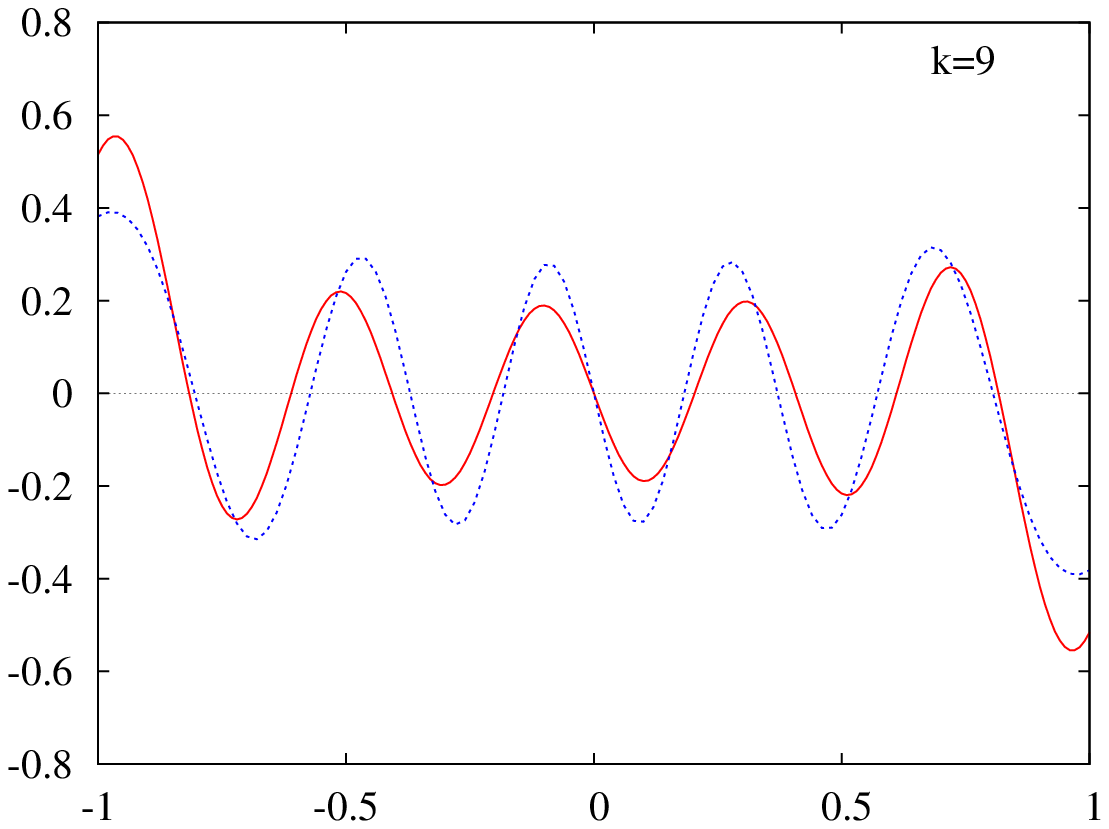}
\includegraphics[width=0.3\textwidth]{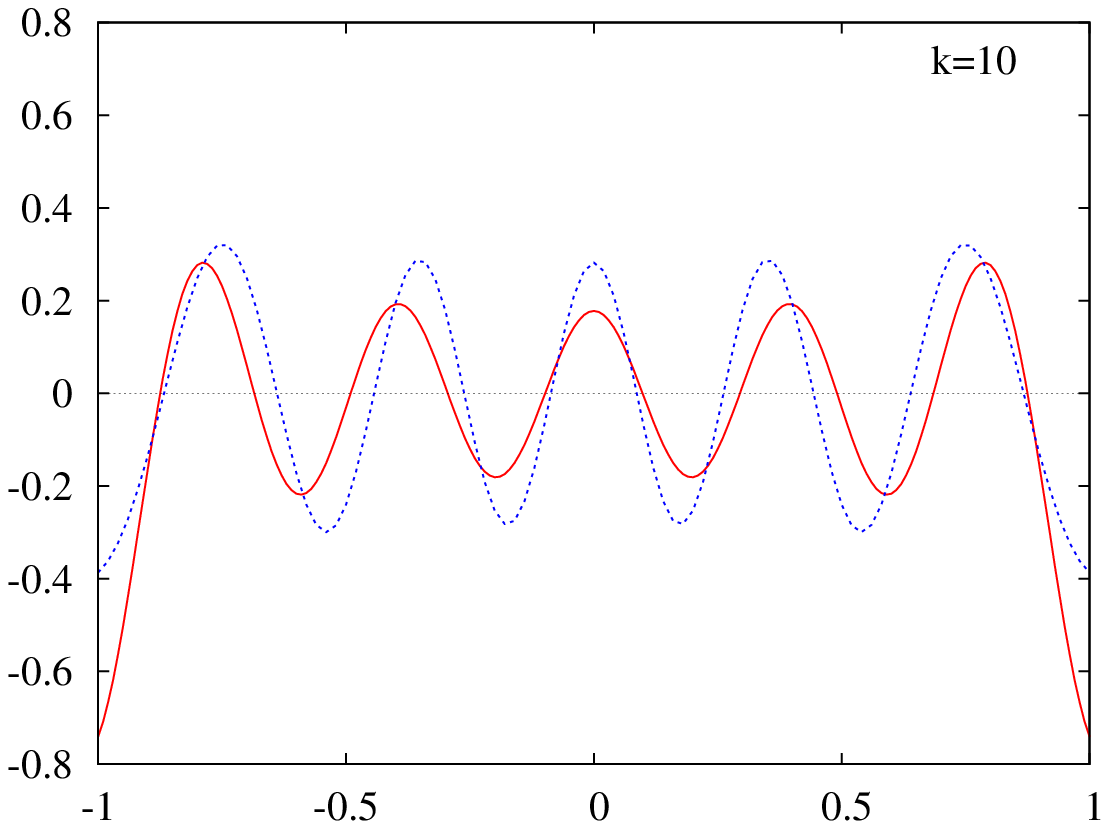}
\includegraphics[width=0.3\textwidth]{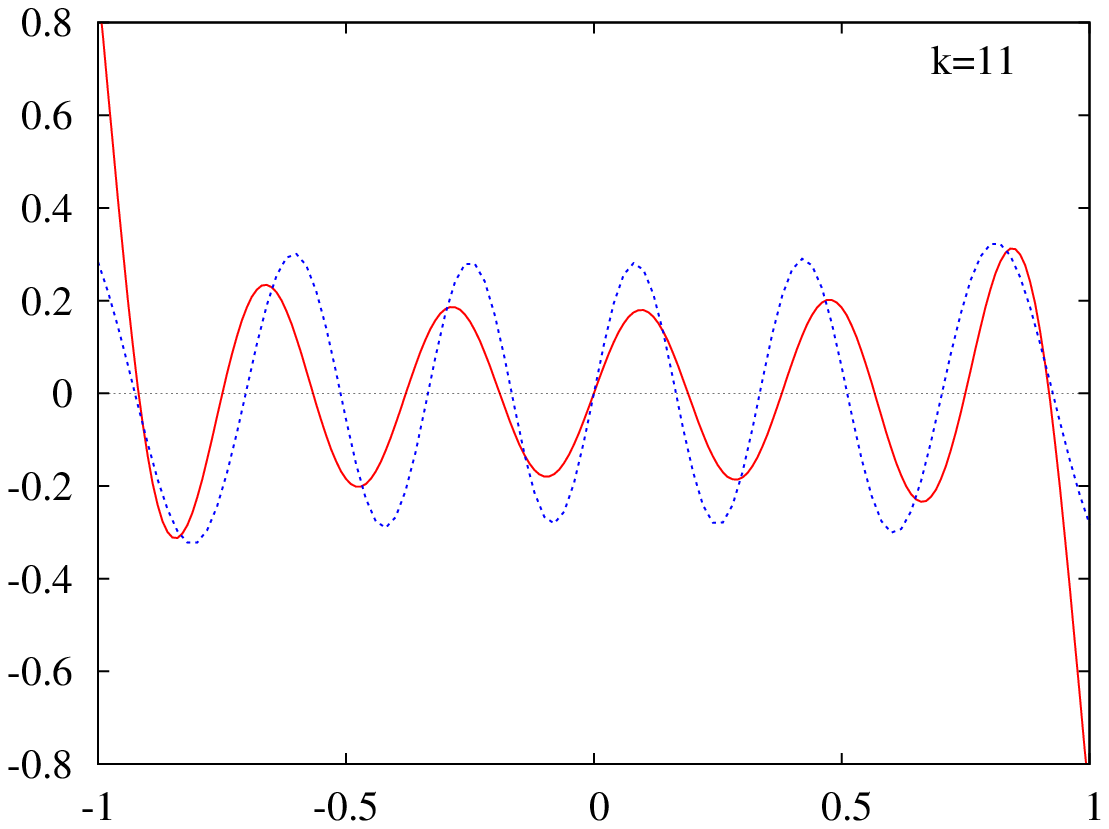}
\includegraphics[width=0.3\textwidth]{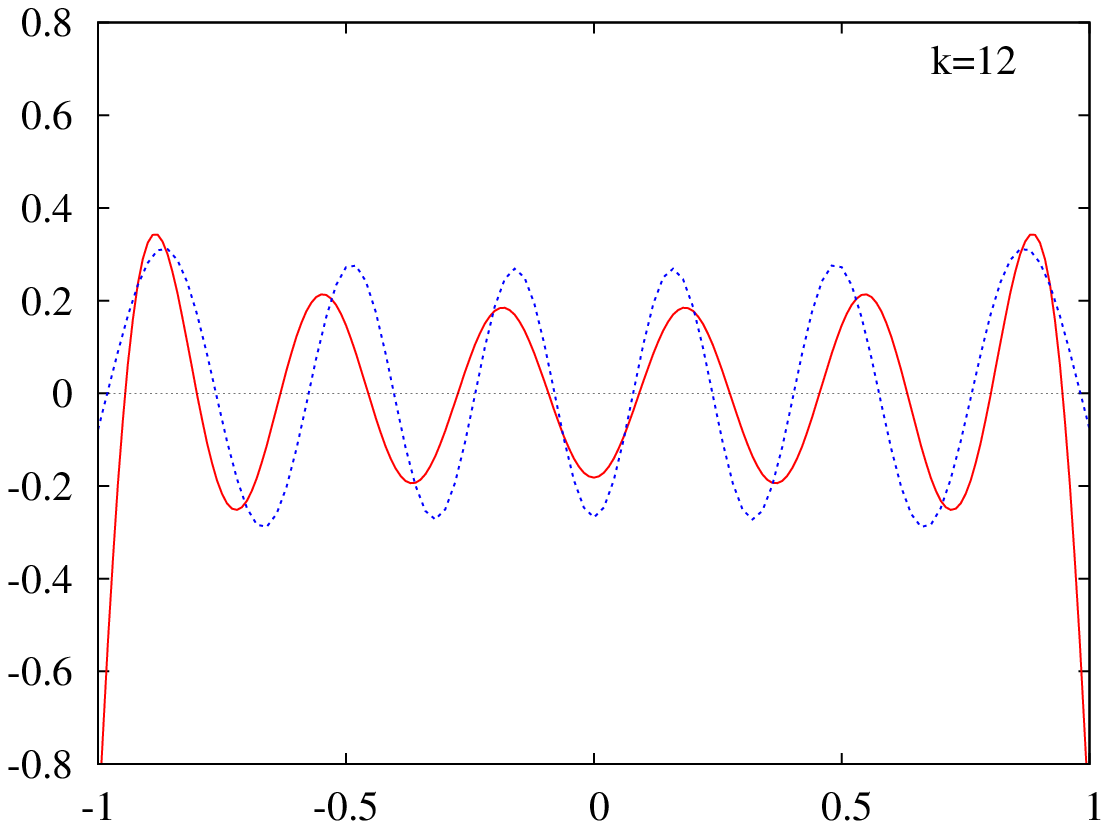}
\includegraphics[width=0.3\textwidth]{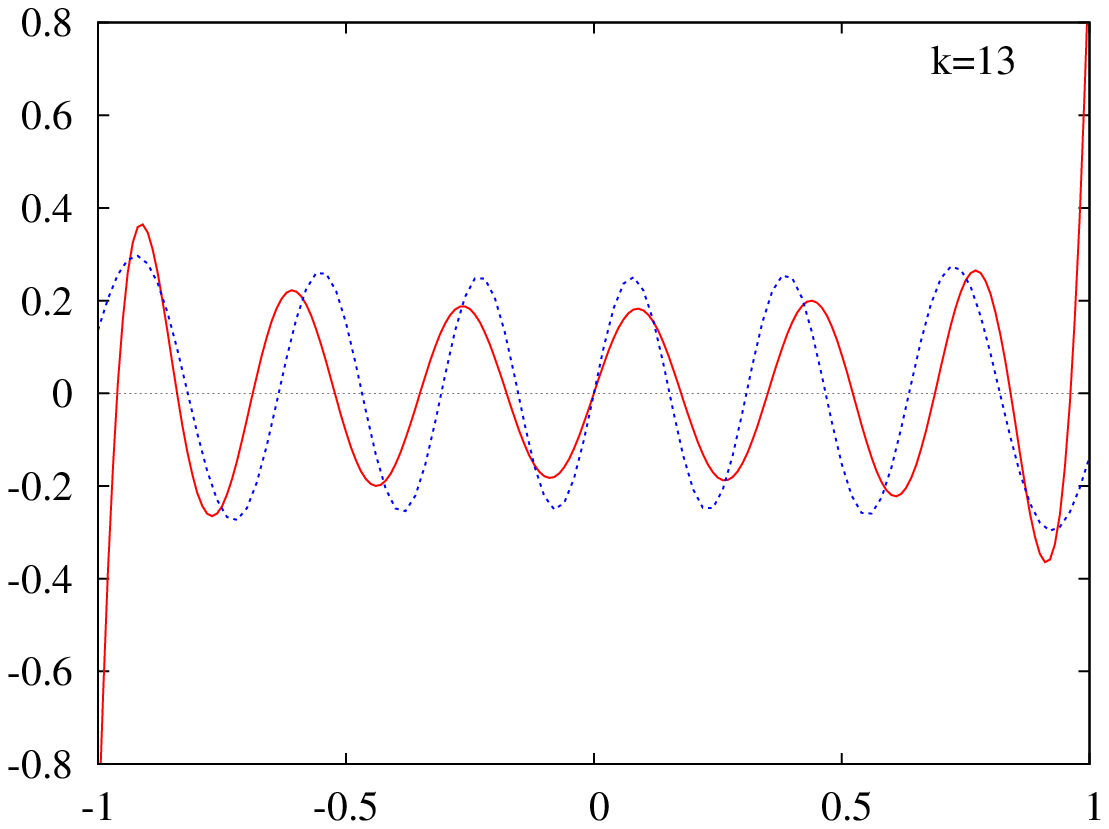}
\includegraphics[width=0.3\textwidth]{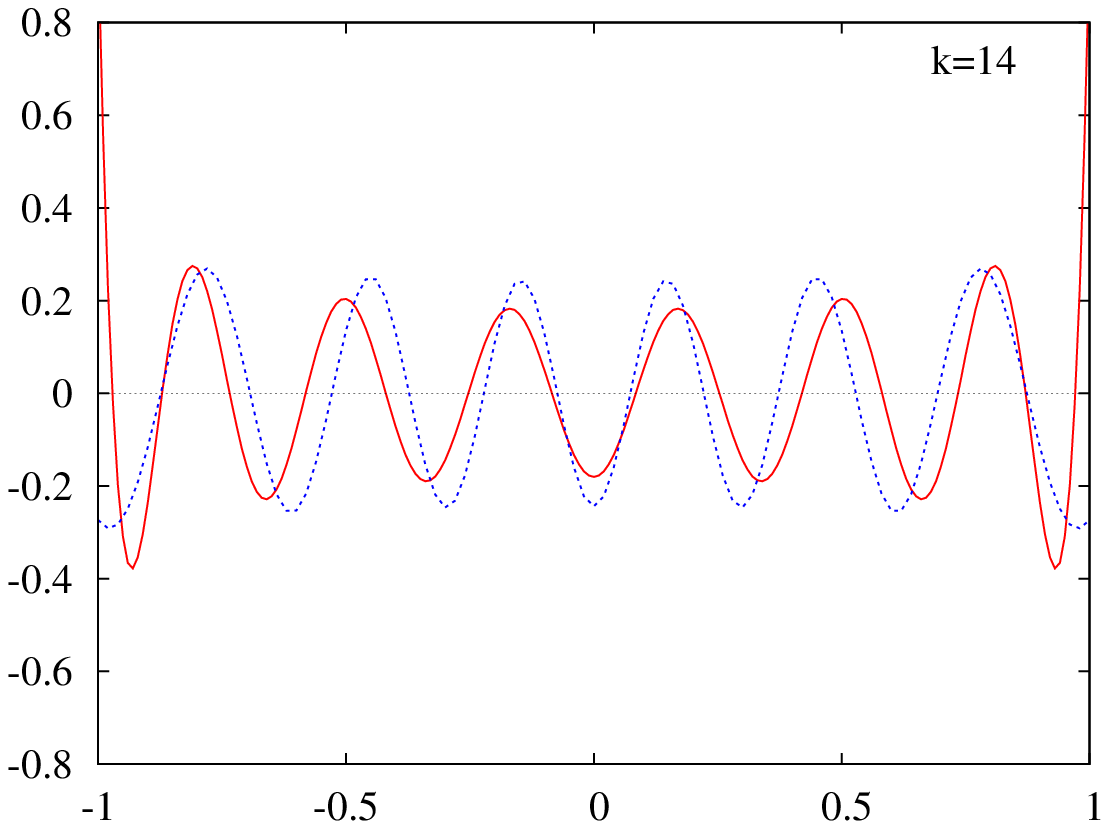}
\caption{The first 15 spheroidal functions $S_{0k}(c,y)$ for $c=5\pi$ (full lines, red)
         and, for comparison, the oscillator functions $u_k(y)$ (dotted lines, blue).}
\label{fig:spheroidS}
\end{figure}

 A qualitative picture of the spheroidal functions can be obtained by observing that for small
$y$ the operator $D$ reduces to 
\begin{equation}
D = - \frac{\dd^2}{\dd y^2} + c^2y^2 
\label{diff_op1}
\end{equation}
and thus to the Hamiltonian of the harmonic oscillator with frequency $\omega=c$. The
oscillator functions 
\begin{equation}
u_k(y) = A_k \, e^{-\frac{1}{2}cy^2} H_k(\sqrt{c} y) 
\label{osc_funct}
\end{equation}
with the Hermite polynomials $H_k$
have a spatial extension $y_k \simeq \sqrt{(2k+1)/c}$ (using the classical turning point)
and therefore "fit" into the subsystem as long as $2k+1 < c$. Under this condition the 
exact spheroidal functions resemble them closely, in particular those concentrated fully
in the interior. A picture gallery is shown in Fig. \ref{fig:spheroidS}, where we plotted 
the first 15 spheroidal and oscillator functions for $c=5\pi$ ($\bar N =10)$. Both are
normalized to $2/(2k+1)$ in $(-1,1)$. The function $S_{0k}$ has exactly $k$ 
zeros there and is alternatingly even and odd. One sees that up to $k \sim 7$ the differences 
are rather small. As $k$ increases further, the oscillator functions deviate more and have
zeros outside the subsystem, while the oscillation of $S_{0k}$ becomes faster near the boundaries. 
For large $c$, it is logarithmic up to a distance $1/\sqrt c$ from $\pm 1$, i.e. besides $y$ 
the variable $\ln[(1+y)/(1-y)]$ enters, as also found in a continuum approximation for a 
half-filled lattice system \cite{Peschel04}. The nature of the eigenfunctions is reflected 
directly in the occupation numbers $\zeta_k$ which are close to 1 for small $k$ and close to 
zero for large $k$. As discussed below, the transition takes place at $k \sim \bar N$.

\begin{figure}[htb]
\center
\includegraphics[width=0.3\textwidth]{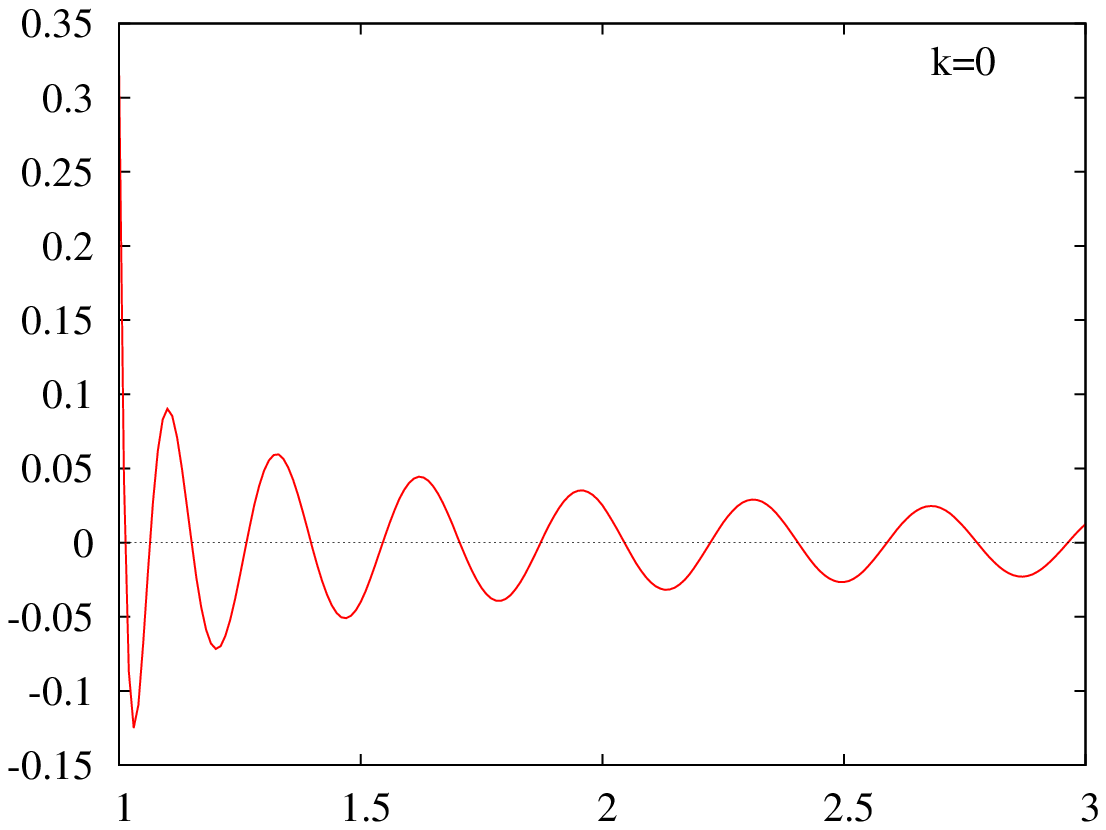}
\includegraphics[width=0.3\textwidth]{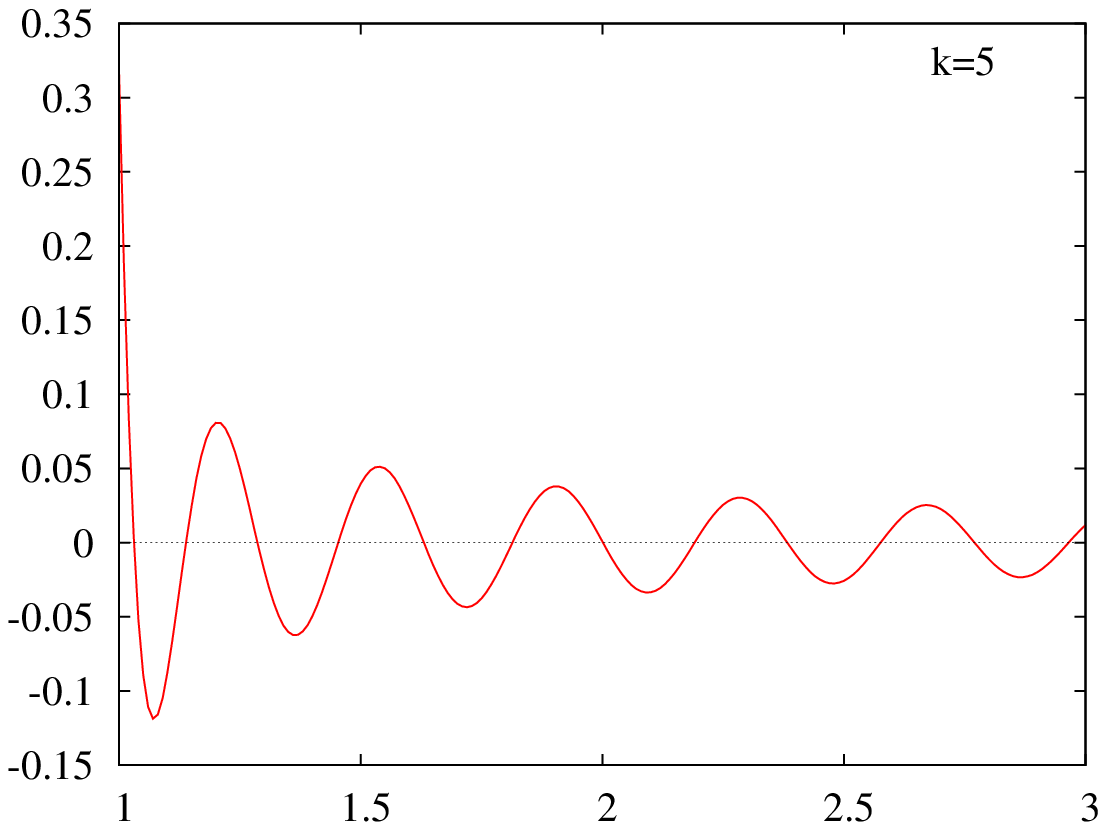}
\includegraphics[width=0.3\textwidth]{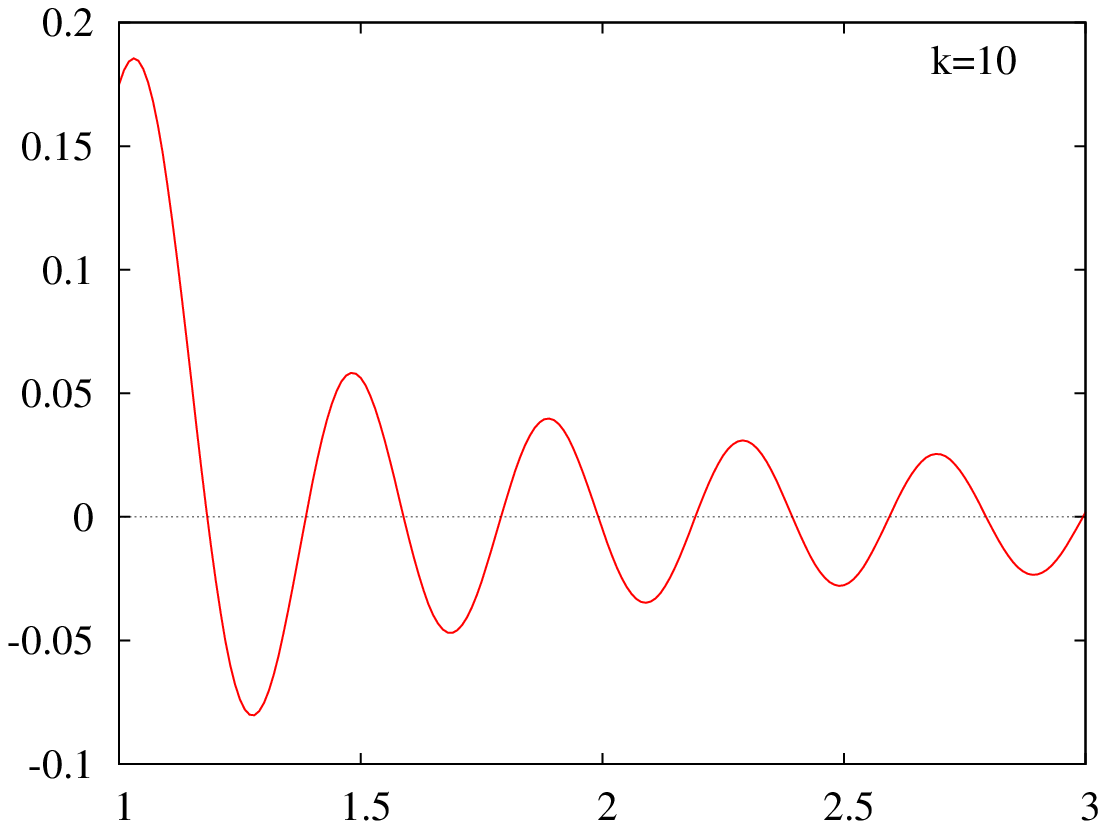}
\includegraphics[width=0.3\textwidth]{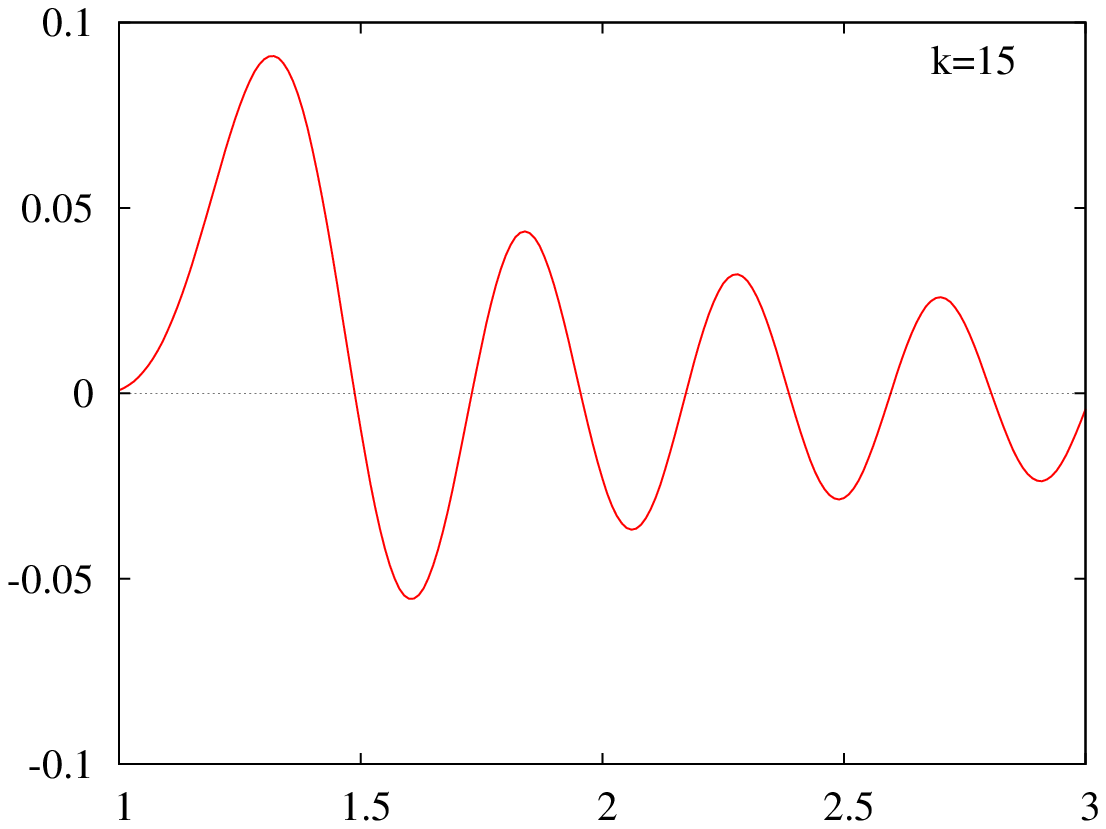}
\includegraphics[width=0.3\textwidth]{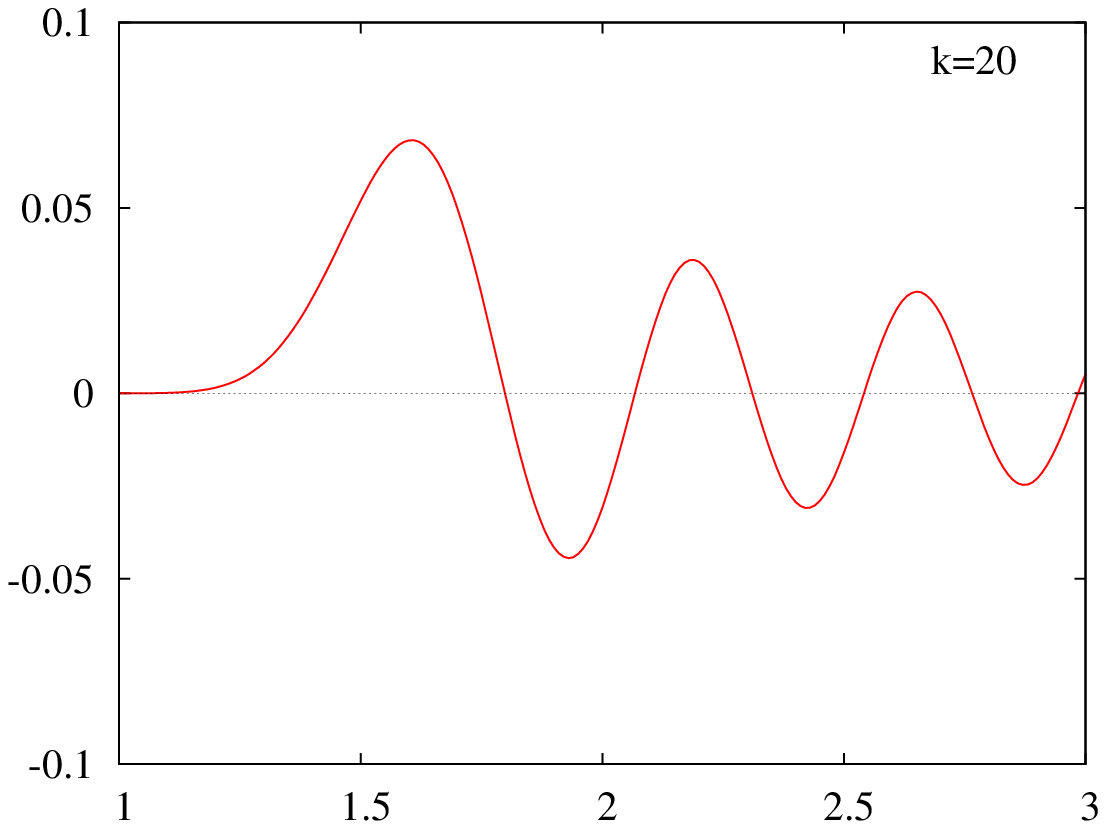}
\includegraphics[width=0.3\textwidth]{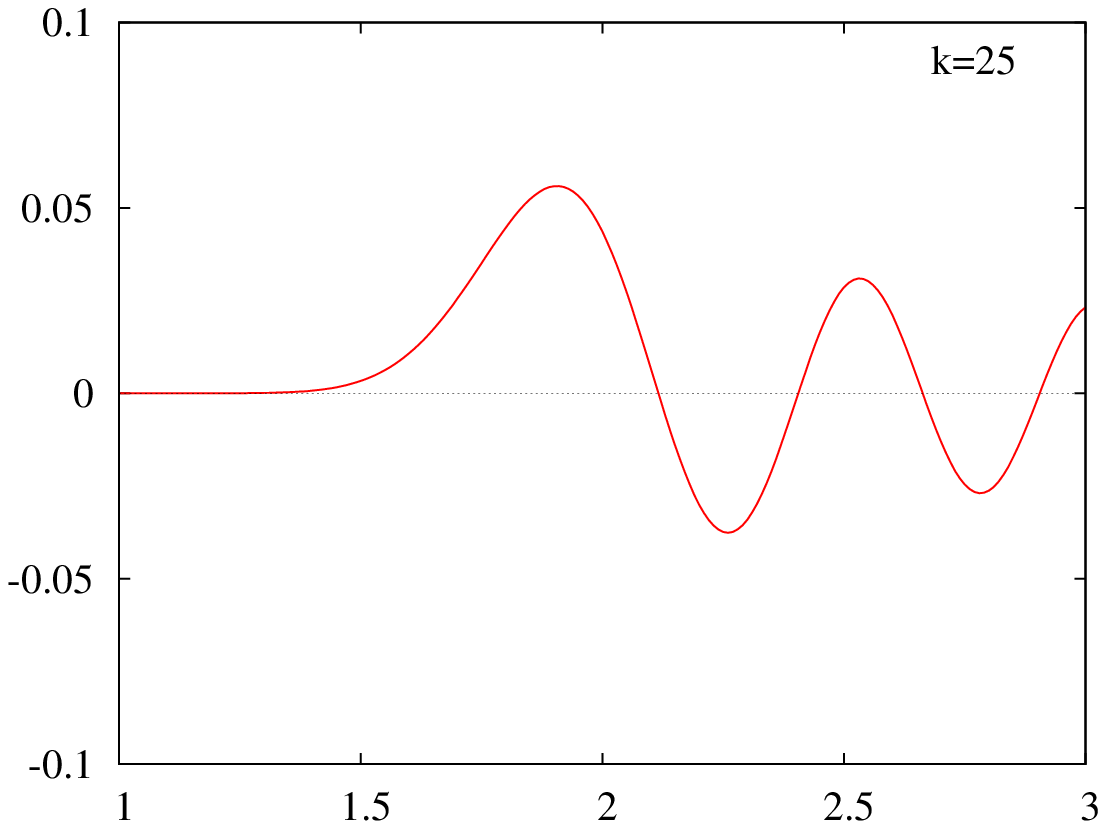}
\includegraphics[width=0.3\textwidth]{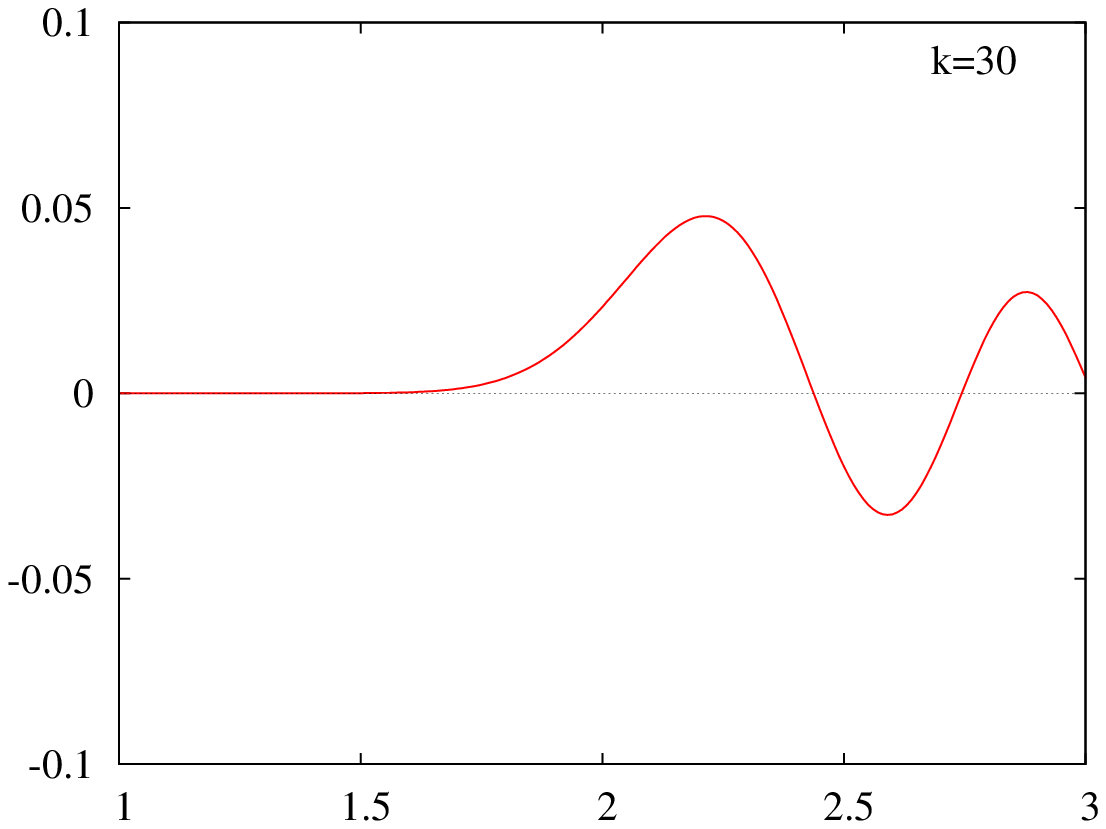}
\includegraphics[width=0.3\textwidth]{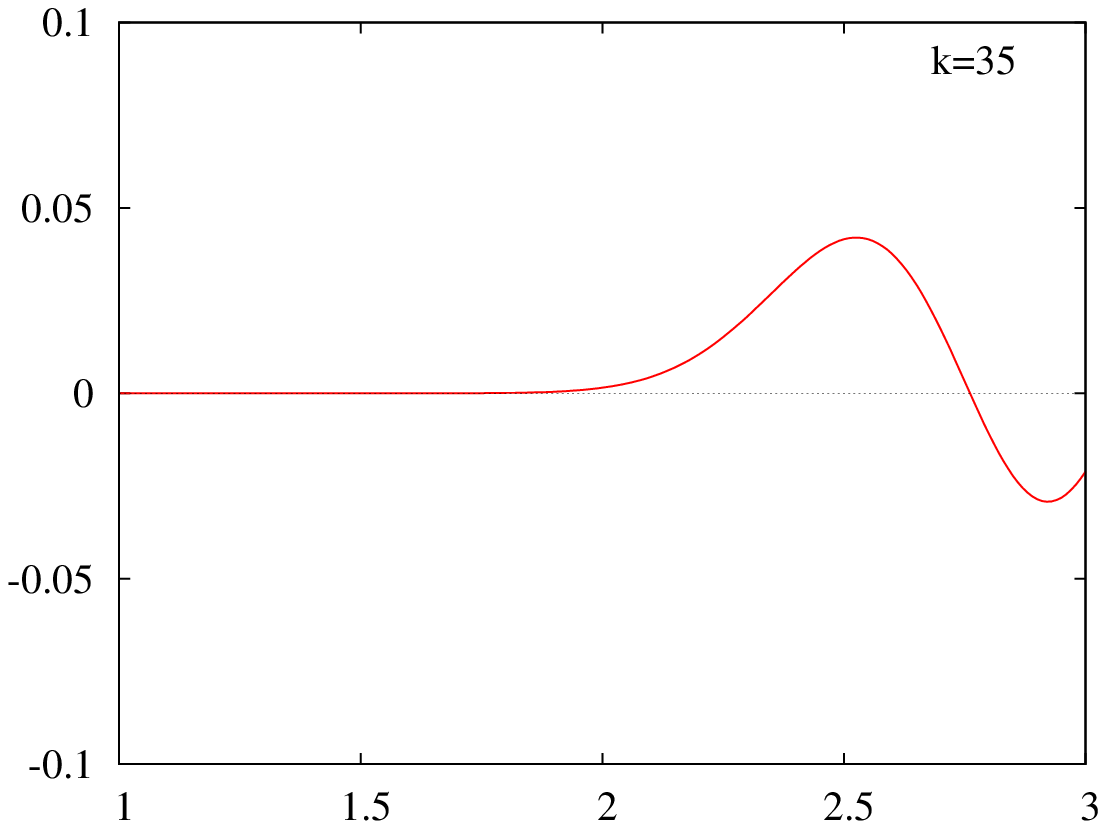}
\includegraphics[width=0.3\textwidth]{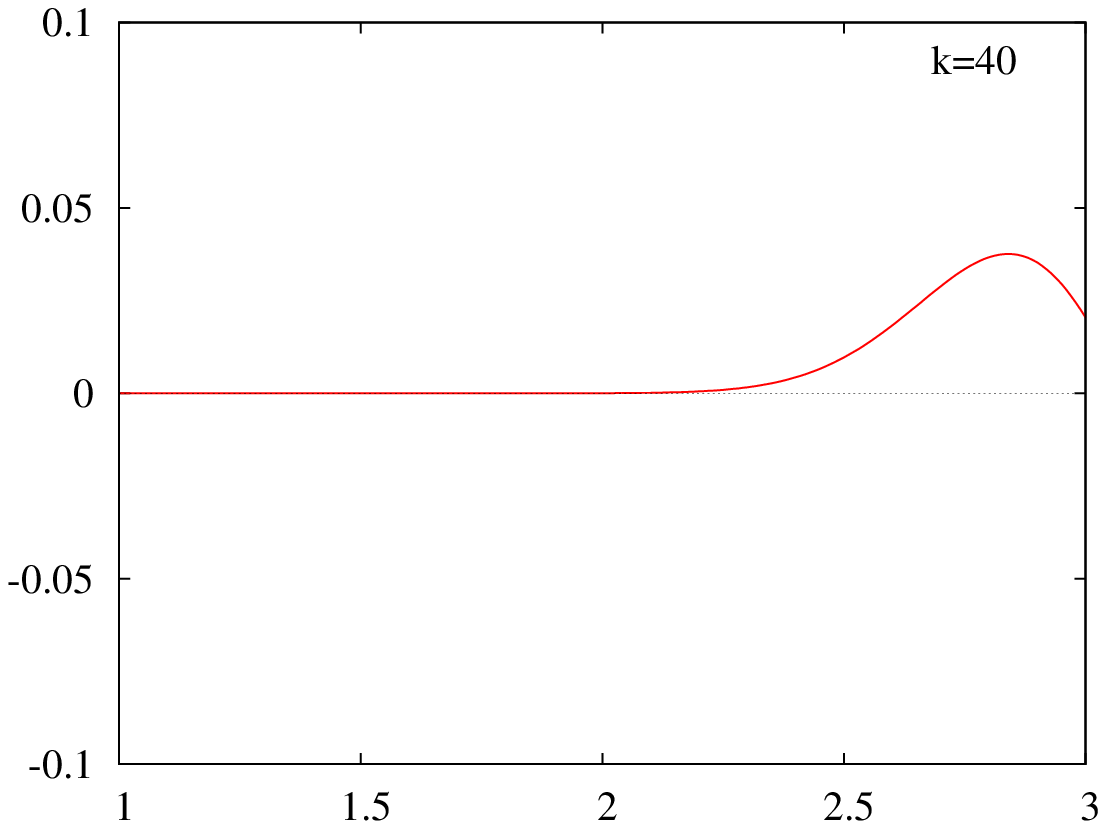}
\caption{Spheroidal functions $R_{0k}(c,y)$ for $c=5\pi$, $1 \le y \le 3$  and various k.}
\label{fig:spheroidR}
\end{figure}

If one continues $\psi_k$ to $y^2 > 1$, one finds the eigenfunctions in the remainder $R$
corresponding to the eigenvalue $1-\zeta_k$. These are a kind of mirror image of those in $S$.
For small $k$, they are oscillating rapidly near the boundary but have very small amplitudes
(the phenomenon has been termed superoscillation \cite{Kempf/Ferreira04}), while for large $k$ 
the behaviour is smooth there and only later an oscillation sets in. This is illustrated in Fig. 
\ref{fig:spheroidR} where the $R_{0k}$ are shown for a number of $k$ values up to $k=40$. Note 
that due to the normalization conventions the functions $S_{0k}$ and $R_{0k}$ do not match at 
$y=1$ The form for $|y| \gg 1$ can be obtained from $D$ by writing $\psi(y) = \chi(y)/y$ which 
leads to the operator
\begin{equation}
\hat D = y \left (\frac{\dd^2}{\dd y^2} + c^2 \right)
\label{diff_op2}
\end{equation}
and gives approximately wave-like solutions $\chi(y) \sim \cos (cy+\alpha) = 
\cos (q_Fx+\alpha)$. This can be seen clearly in the figure.

%
%

 Turning to the eigenvalues $\zeta_k$ (usually called $\lambda_k$ in the literature), they can
be obtained from the relation \cite{Flammer57} 
\begin{equation}
\int_{-1}^{1} \dd y \, \ee^{icyz} S_{0k}(c,y) = 2 i^n R_{0k}(c,1) S_{0k}(c,z)
\label{flammer}
\end{equation}
which is, up to a factor, the eigenvalue equation of the integral operator $\tilde K$. 
This gives the general formula
\begin{equation}
\zeta_k = |\mu_k|^2 = \frac{2c}{\pi}  \left[ R_{0k}(c,1) \right]^2
\label{zeta_gen}
\end{equation}
which can be evaluated numerically.
To obtain analytical expressions for large $c$, Slepian considered the solutions of 
(\ref{diff_eigen}) in the various regions of $y$ and connected them in the domains of overlap 
\cite{Slepian65}. While he considered small and large $k$ separately, des Cloizeaux and Mehta
using WKB formulae could later obtain a single expression \cite{DesClo/Mehta73} and 
Landau and Widom gave a mathematical proof \cite{Landau/Widom80}. The procedure is 
rather tedious, but in the end a relatively simple result for $\varepsilon_k$ emerges if 
$k \simeq \bar N$, c.f. (1.37), (1.38) in \cite{Slepian65}. Namely, it is given by the root 
of smallest absolute value of the equation
\begin{equation}
c + \frac{\varepsilon_k}{\pi} \ln(2\sqrt{c})-\varphi(\frac{\varepsilon_k}{2\pi}) =
\frac{\pi}{2}(k-\frac{1}{2})
\label{epsilon_as1}
\end{equation}
where $\varphi(z)=\mathrm{arg}\Gamma(1/2+iz)$ and $\Gamma(z)$ is the gamma function. This
has solutions $|\varepsilon_k| \ll 1$ which are obtained by expanding $\varphi(z)$
\begin{equation}
\frac{\varepsilon_k}{2\pi} \left(\ln(4c)-\varphi'(0)\right) =
\frac{\pi}{2}(k- 1/2 - \bar N)
\label{epsilon_as2}
\end{equation}
or, with $c$ inserted,
\begin{equation}
\varepsilon_k = \frac{\pi^2(k- 1/2 - \bar N)}{\ln(2\pi\bar N)-\psi(1/2)} 
\label{epsilon_as3}
\end{equation}
where $\psi(1/2)=-\gamma-2\ln2=-1.963...$, $\gamma$ is Euler's constant and $\psi(z)$ 
denotes the digamma function. Thus $\varepsilon_k$ goes through zero ($\zeta_k$ through 1/2) 
for $k \simeq \bar N \gg 1$ and the slope vanishes logarithmically with $c$ or $\bar N$. 
A similar formula was found in \cite{Peschel04} for a half-filled lattice system. For larger
$\varepsilon_k$ one really has to solve (\ref{epsilon_as1}). It is simpler, however, to view
it as a relation $k=k(\varepsilon)$ which leads to a density of states 
$n(\varepsilon)=dk/d\varepsilon$ given by
\begin{equation}
n(\varepsilon)=\frac{1}{\pi^2}\left [\ln(4c)-\varphi'(\frac{\varepsilon}{2\pi})\right]
\label{density_states1}
\end{equation}
where explicitly 
\begin{equation}
\varphi'(z) = \frac{1}{2}[\, \psi(\frac{1}{2}+iz)+\psi(\frac{1}{2}-iz)\, ]
\label{density_states2}
\end{equation}
This is the result found and plotted in \cite{Suesstrunk/Ivanov12} and implicit already
in \cite{Jin/Korepin04}. Inserting it into the expression for the entanglement entropy,
one finds the leading logarithmic term and the first correction. We will come back to it
in Section 5.

%
%

\section{Finite continuous system}

  Consider now a finite system in the form of a ring with circumference $L$. The momenta
are quantized according to $q=2\pi n/L$ with integer $n$ and the Fermi momentum is
written as $q_F=2\pi m/L$. This gives a total particle number $N=2m+1$ and a correlation 
matrix
\begin{equation}
C(x-x') = \frac{1}{L} \sum_{n=-m}^{m}\, e^{-i\,2\pi(x-x')n/L} \\
        = \frac{1}{L} \frac{\sin{(N\pi/L)(x-x')}}{\sin{(\pi/L)(x-x')}}
\label{corr_cont_fin1}
\end{equation}
With the same subsystem as before, $-\ell/2 \le x \le \ell/2$, the eigenvalue equation
(\ref{corr_cont2}) retains its form. However, here it is preferable to define the reduced 
variable as $z=x/L$ such that $|z| \le 1/2$. Then the equation becomes
\begin{equation}
\int_{-W}^{W} dz' \, K'(z-z') \psi_k(z') = \zeta_k \,\psi_k(z) 
\label{corr_cont_fin2}
\end{equation}
with $W=\ell/2L < 1/2$ and the \emph{modified sine kernel}
\begin{equation}
K'(z-z') = \frac{\sin{\pi N (z-z')}}{\sin{\pi(z-z')}}
\label{mod_sine_kern}
\end{equation}
The overlap matrix is 
\begin{equation}
A(q-q') = \int_{-\ell/2}^{\ell/2} \frac{dx}{L} \, e^{-i(q-q')x} =  
           \frac{\sin (q-q')\ell/2}{(q-q')L/2} 
\label{overlap_cont_fin1}
\end{equation}
but since the momenta are discrete, one can write it as
\begin{equation}
A_{nn'} = \frac{\sin{(\pi\ell/L)(n-n')}}{\pi(n-n')}
\label{overlap_cont_fin2}
\end{equation}
and the eigenvalue equation is a genuine matrix equation with a Toeplitz matrix.

The problem is now in exactly the form as studied by Slepian in 1978 \cite{Slepian78}. In
particular, (\ref{corr_cont_fin2}) is his equation (10) and the $\psi_k$ are his functions
$U_k$. There are only $N$ of them with non-zero eigenvalue (corresponding 
to the $N$ eigenvalues of the overlap matrix) and he calls them 
\emph{discrete prolate spheroidal wave functions}. The treatment of the infinite case can be 
repeated, because there is again a commuting differential operator, namely
\begin{equation}
D' = - \frac{1}{4\pi^2 }\frac{\dd}{\dd z} (\cos{2\pi z}-\cos{2\pi W}) \frac{\dd}{\dd z} 
     - \frac{1}{4} (N^2-1)\cos{2\pi z}
\label{diff_op_fin}
\end{equation}
One sees that, in comparison with (\ref{diff_op}), the powers have been replaced by
cosine functions, which reflect the ring geometry. The first term is a kinetic energy
with different sign in $S$ and $R$, as in the infinite case. The second term is a potential
with a minimum at the origin. In the limit of small $z$ and $W$ one can expand the cosine
functions and recovers the operator $D$ of the infinite case. The finite size is then no
longer visible. The same holds for the integral operator $K'$ which reduces to $K$. The $N$
lowest eigenfunctions of the differential operator give the solution both in the subsystem and 
in the remainder, and determining the respective norms allows to calculate the eigenvalues 
$\zeta_k$. Pictures of the functions for $N=4,5$ and $W=0.2,0.4$ are shown in \cite{Slepian78}
and one sees the similarity to the oscillator functions in particular for the lowest state $k=0$.

  The analysis of the eigenfunctions of $D'$ for large $N$ leads to formulae for 
$\varepsilon_k$ which are very similar to those of the infinite system, see equs. (53), (60)
and (61) in \cite{Slepian78}. Instead of (\ref{epsilon_as1}) one finds the equation
\begin{equation}
\pi N W + \frac{\varepsilon_k}{2\pi} \ln(2N\sin{2\pi W})-\varphi(\frac{\varepsilon_k}{2\pi}) =
\frac{\pi}{2}(k-\frac{1}{2})
\label{epsilon_fin_as1}
\end{equation}
and in the linear region this gives after inserting $W$
\begin{equation}
\varepsilon_k = \frac{\pi^2(k- 1/2 - \bar N)}{\ln(2N \sin{(\pi\ell/L)})-\psi(1/2)} 
\label{epsilon_fin_as3}
\end{equation}
For $L \rightarrow \infty$, keeping $N/L=\bar n$ constant, this reduces to (\ref{epsilon_as3}). 

The existence of the commuting operator $D'$ also has consequences for the other eigenvalue 
problem. This will be discussed in the following section.

\section{Infinite lattice system}

  In the entanglement investigations, this is the most studied case. Let the system be the 
limit of a ring with a total of $M$ sites. The correlation matrix is 
\begin{equation}
C_{nn'} = \int_{-q_F}^{q_F} \frac{dq}{2\pi} \, e^{-iq(n-n')} =  
           \frac{\sin q_F(n-n')}{\pi (n-n')} 
\label{corr_latt1}
\end{equation}
which corresponds to (\ref{corr_cont1}) but with discrete sites. We will choose the
subsystem here as the $L$ sites $n=1,2,...L$. Then the eigenvalue equation is
\begin{equation}
\sum_{n'=1}^{L}  C_{nn'} \,\varphi_k(n') = \zeta_k \,\varphi_k(n)  
\label{corr_latt2}
\end{equation}
and for the overlap matrix one finds
\begin{eqnarray}
A(q-q') &=& \frac{1}{M}\sum_{n=1}^{L}e^{-i(q-q')n} \nonumber \\ 
        &=&  e^{-iq(L+1)/2} \,\left (\frac{1}{M} \frac{\sin{(q-q')L/2}}{\sin{(q-q')/2}} 
             \right) \, e^{iq'(L+1)/2}
\label{overlap_latt1}
\end{eqnarray}
The exponential factors reflect the location of the subsystem, but do not affect
the spectrum. The eigenvalue equation can be written
\begin{equation}
\int_{-q_F}^{q_F}  \frac{dq'}{2\pi} \, \frac{\sin{(q-q')L/2}}{\sin{(q-q')/2}} \,
 \tilde\varphi_k(q') = \zeta_k \,\tilde\varphi_k(q),\quad \quad 
 \tilde\varphi_k(q)= e^{iq(L+1)/2} \,\varphi_k(q) 
\label{overlap_latt2}
\end{equation}
With $p=q/2\pi$, $W=q_F/2\pi=\bar n/2$ and $\tilde\varphi_k(2\pi p)= \psi(p)$ this
leads to
\begin{equation}
\int_{-W}^{W} dp' \, K'(p-p') \psi_k(p') = \zeta_k \,\psi_k(p) 
\label{corr_cont_fin22}
\end{equation}
which is (\ref{corr_cont_fin2}) with the substitution $N \rightarrow L$ in the kernel $K'$.

Therefore, as noted in previous work \cite{CMV11a,CMV11b}, the infinite lattice and the finite
continuum problem are actually the same with the role of positions and momenta, i.e.
of correlation and overlap matrices, interchanged. Thus the discrete spheroidal functions 
appear here in momentum space, and the same holds for the commuting differential operator. 
As mentioned above, this operator has a consequence for the other eigenvalue problem
(\ref{corr_latt2}). From (\ref{mutual})  
\begin{equation}
\tilde\varphi_k(q)=\frac{1}{\sqrt{\zeta_k}}\, e^{iq(L+1)/2}  
                    \frac{1}{\sqrt M}\sum_{n=1}^{L}\varphi_k(n)e^{-iqn}
\label{overlap_latt3}
\end{equation}
and inserting this into the eigenvalue equation $D'\tilde\varphi=\theta' \tilde\varphi$
one finds a three-term recursion relation for the $\varphi_k(n)$ which involves only the
sites $n-1,n$ and $n+1$. In other words, the $\varphi_k(n)$ are also the eigenfunctions of a 
tridiagonal matrix $T$ which therefore commutes with $C$ (in the subsystem). Thus the 
relation $[K',D']=0$ has an equivalent $[C,T]=0$ in real space.  

Writing $T$ in the form
\begin{equation}
   T =  
    \left(  \begin{array} {ccccc}
      d_1 & t_1   &  &   & \\  
       t_1 & d_2 & t_2  &  &\\
       & t_2 & d_3 & t_3  &  \\
        & & \ddots & \ddots & \\
        & & & t_{L-1} & d_L
  \end{array}  \right)   
  \label{tridiagonal1}
 \end{equation}
the matrix elements are
 \begin{equation}
  t_n = \frac{1}{2} n(L-n),\;\;\;\;
  d_n = (\frac{L+1}{2}-n)^2 \, \cos{q_F}\; 
  \label{tridiagonal2} 
 \end{equation}
Both coefficients vary parabolically around the center of the system. Viewed as a hopping 
Hamiltonian, $T$ describes an inhomogeneous system where the hopping is largest in the 
middle. The potential energy is also largest there if $q_F > \pi/2$, but smallest if 
$q_F < \pi/2$. For a half-filled system, $q_F = \pi/2$, it vanishes.

\begin{figure}[htb]
\center
\includegraphics[width=0.3\textwidth]{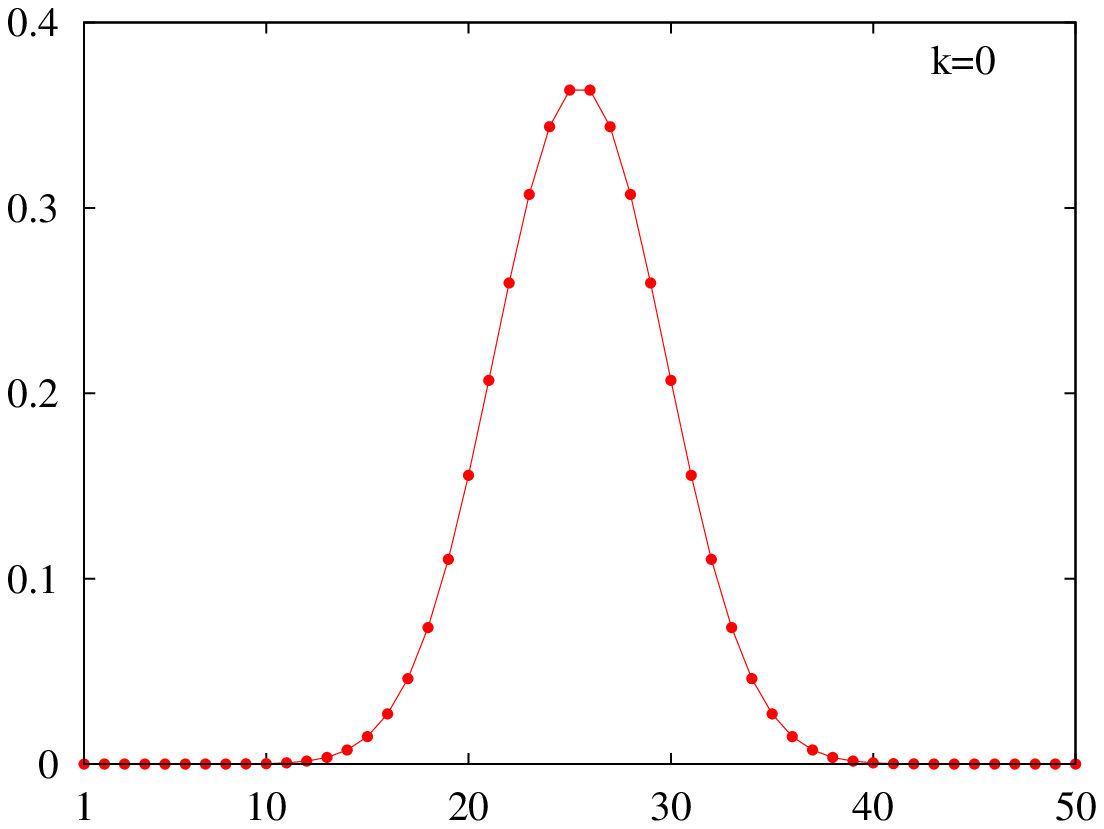}
\includegraphics[width=0.3\textwidth]{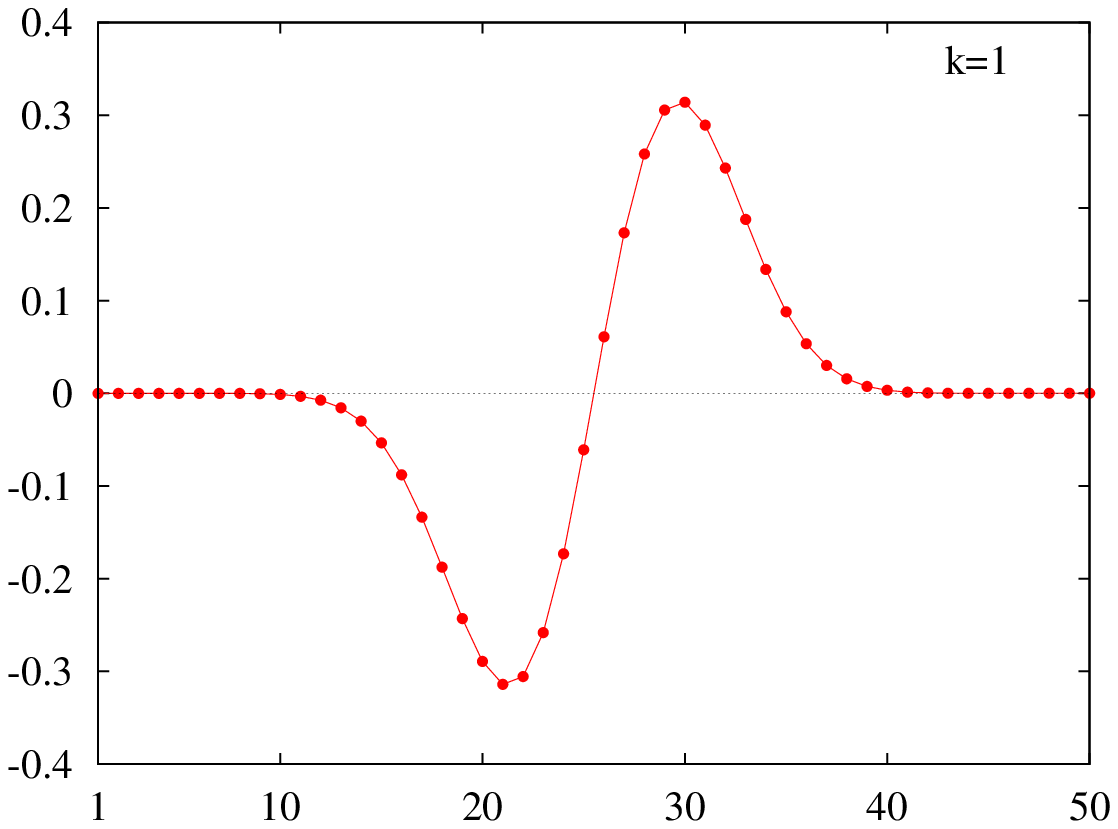}
\includegraphics[width=0.3\textwidth]{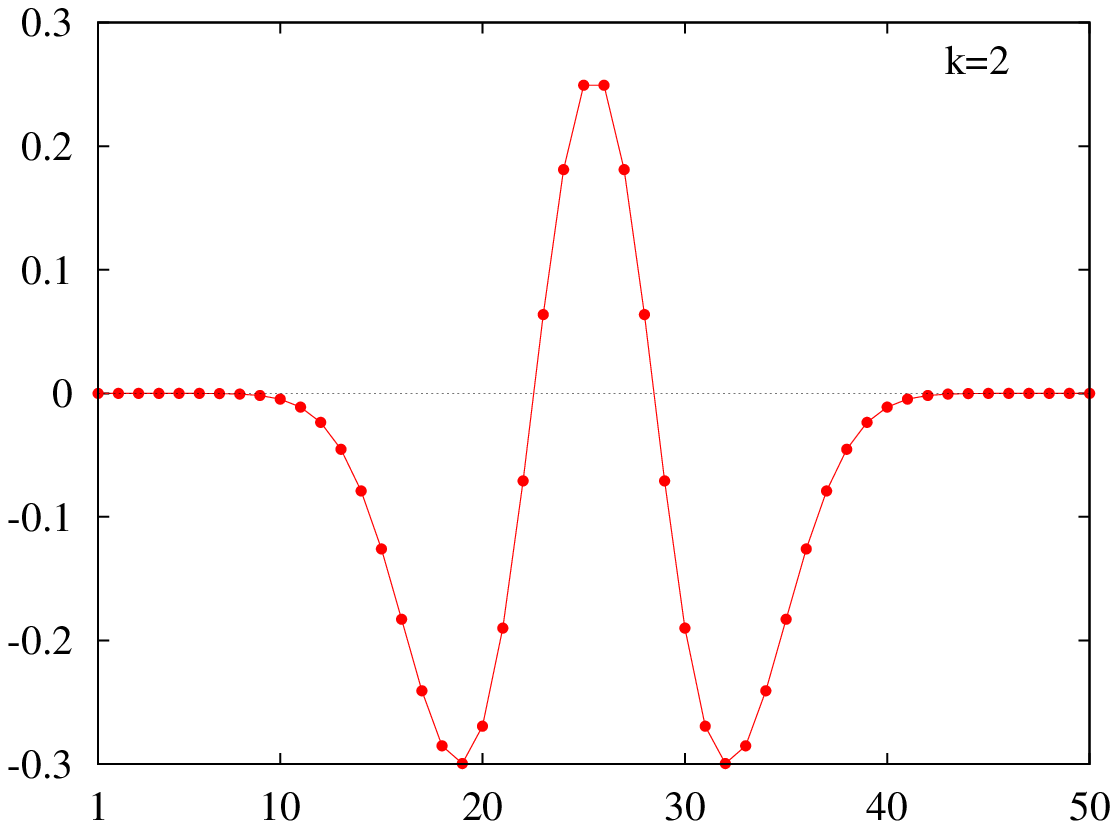}
\includegraphics[width=0.3\textwidth]{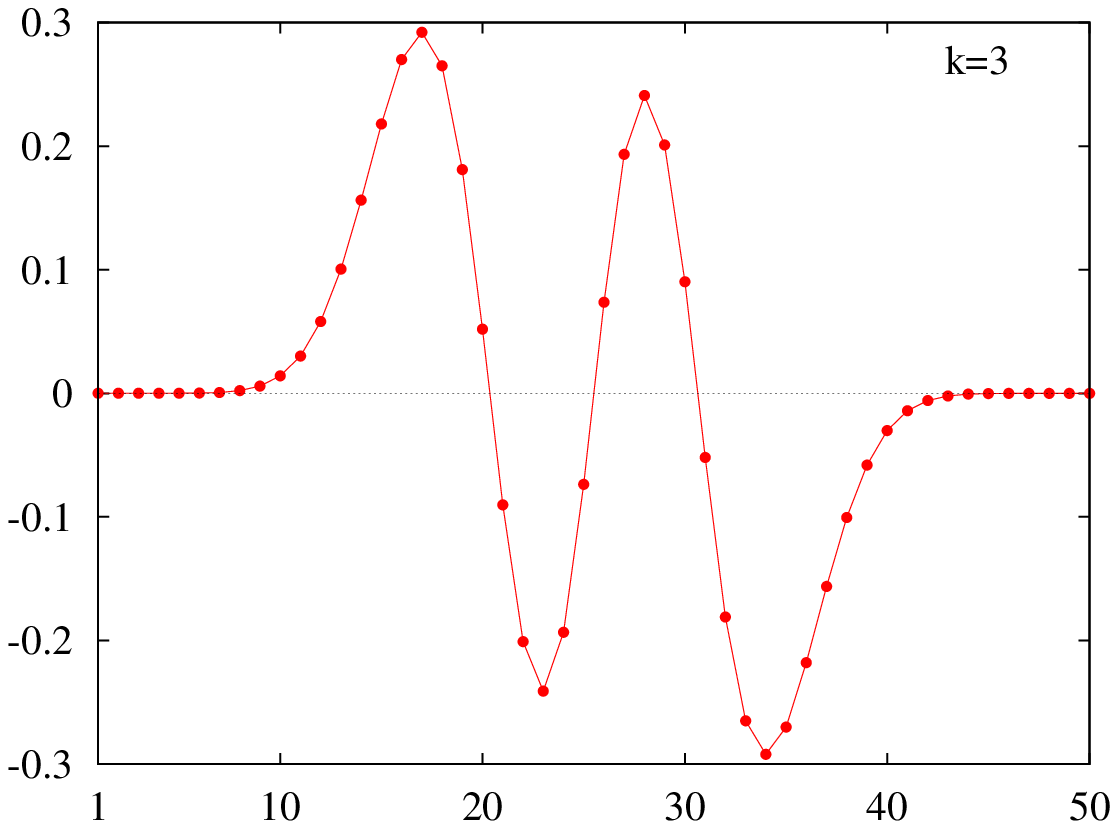}
\includegraphics[width=0.3\textwidth]{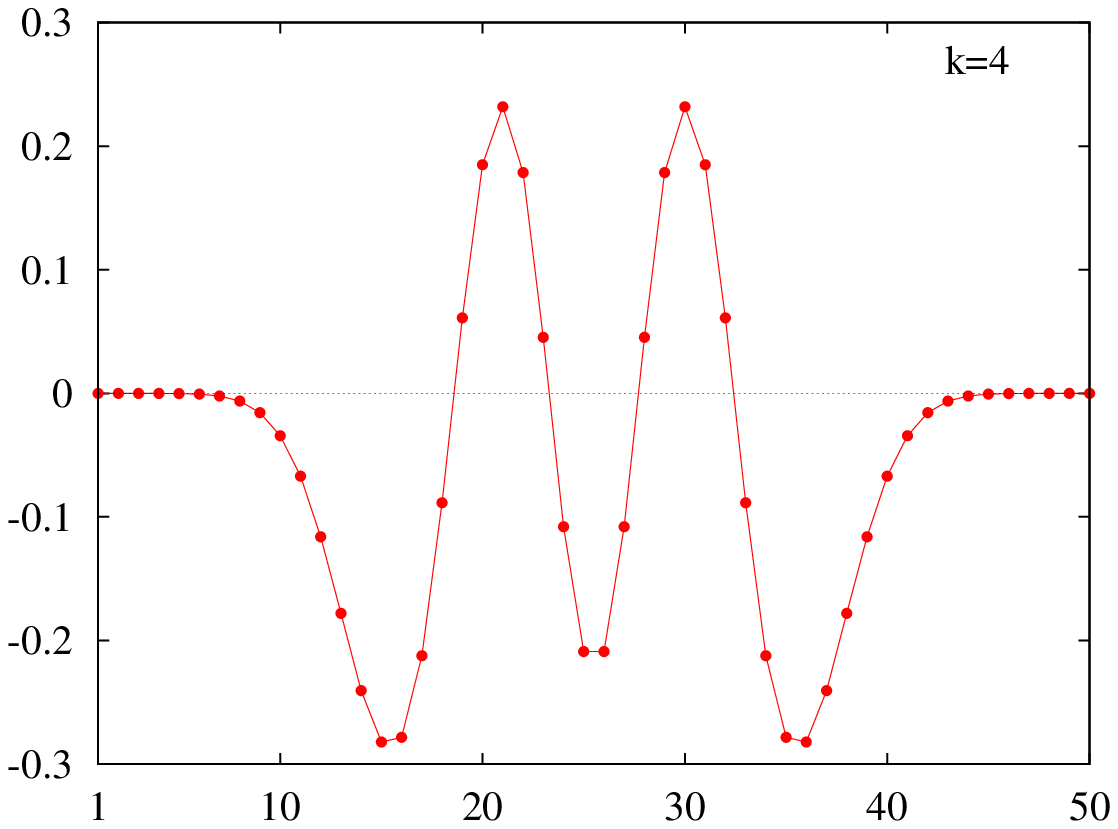}
\includegraphics[width=0.3\textwidth]{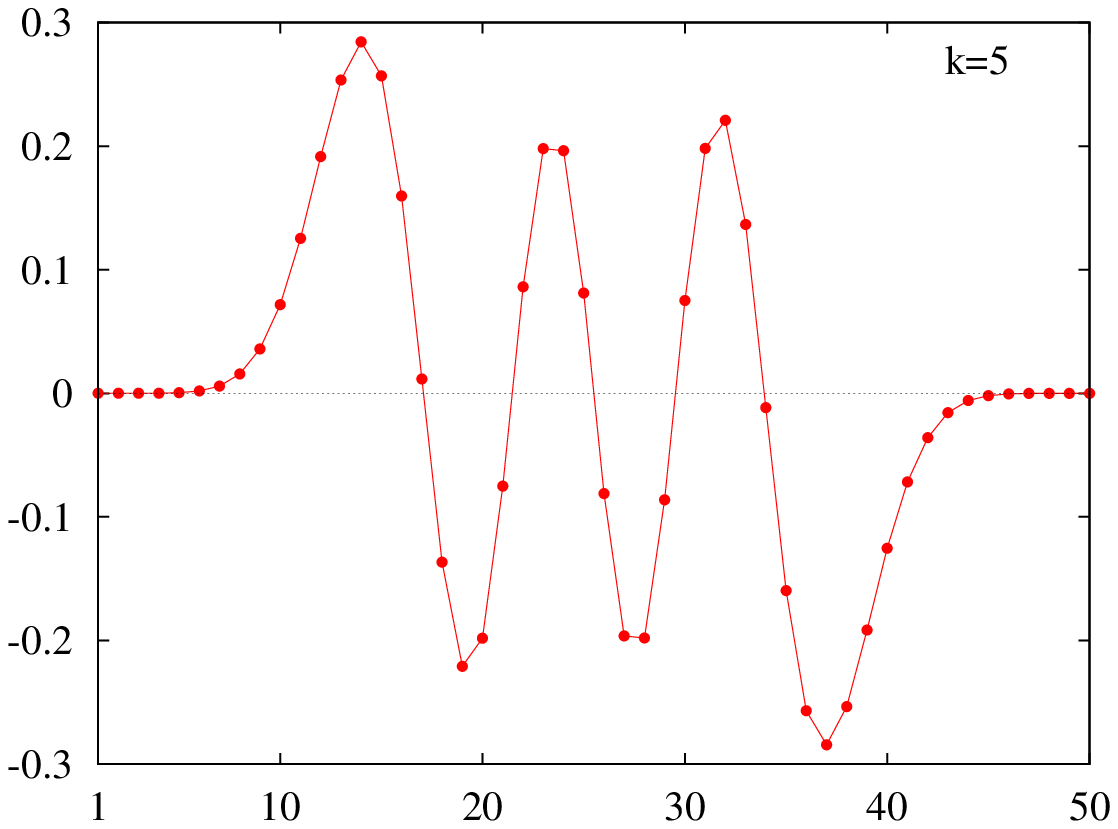}
\includegraphics[width=0.3\textwidth]{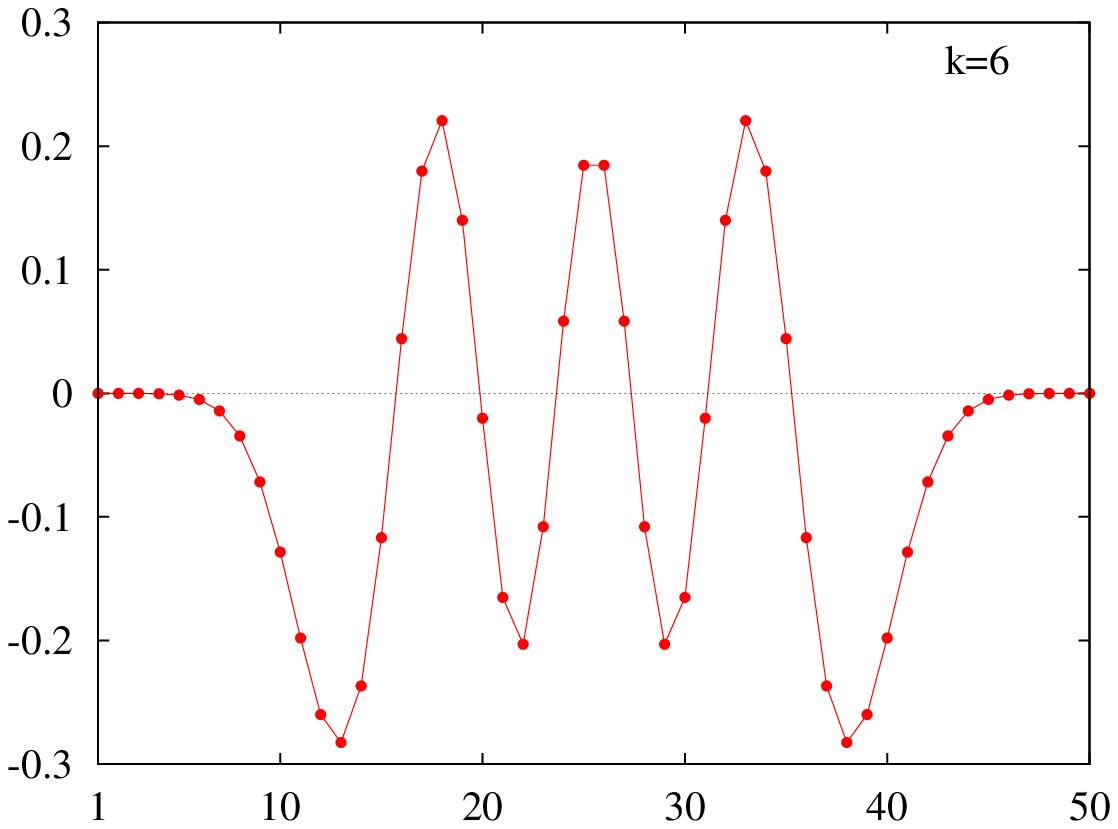}
\includegraphics[width=0.3\textwidth]{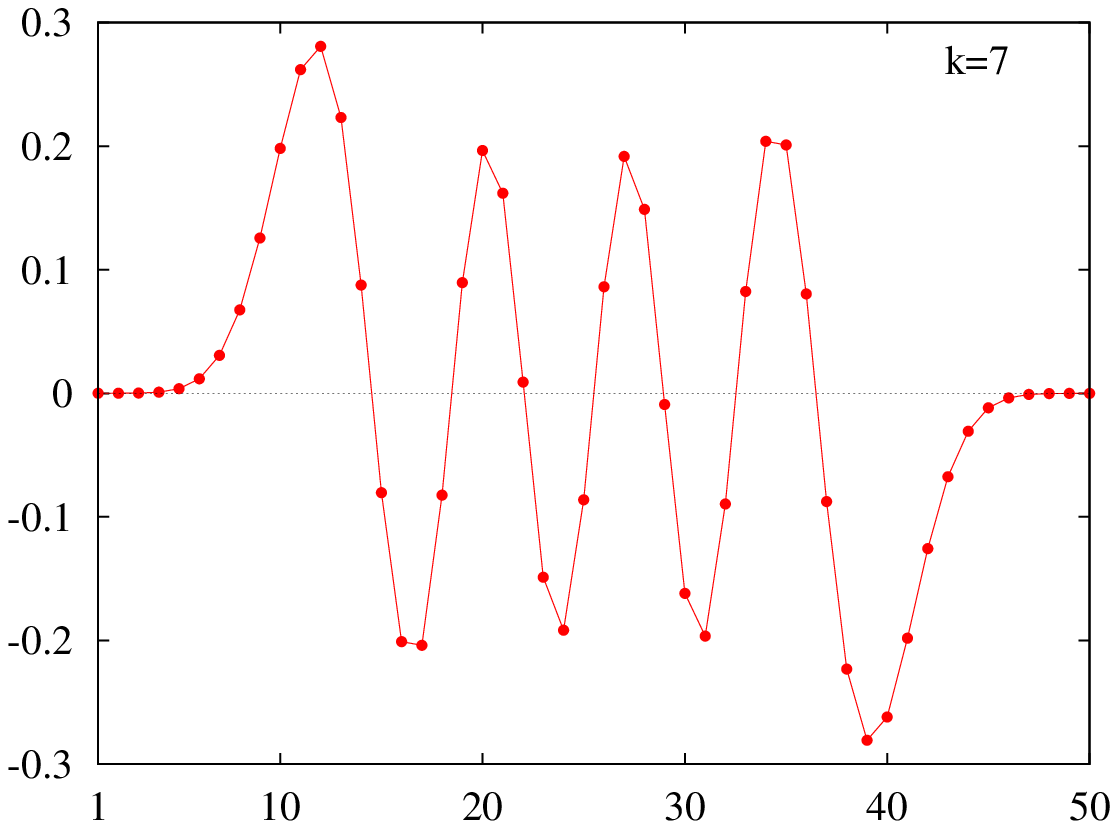}
\includegraphics[width=0.3\textwidth]{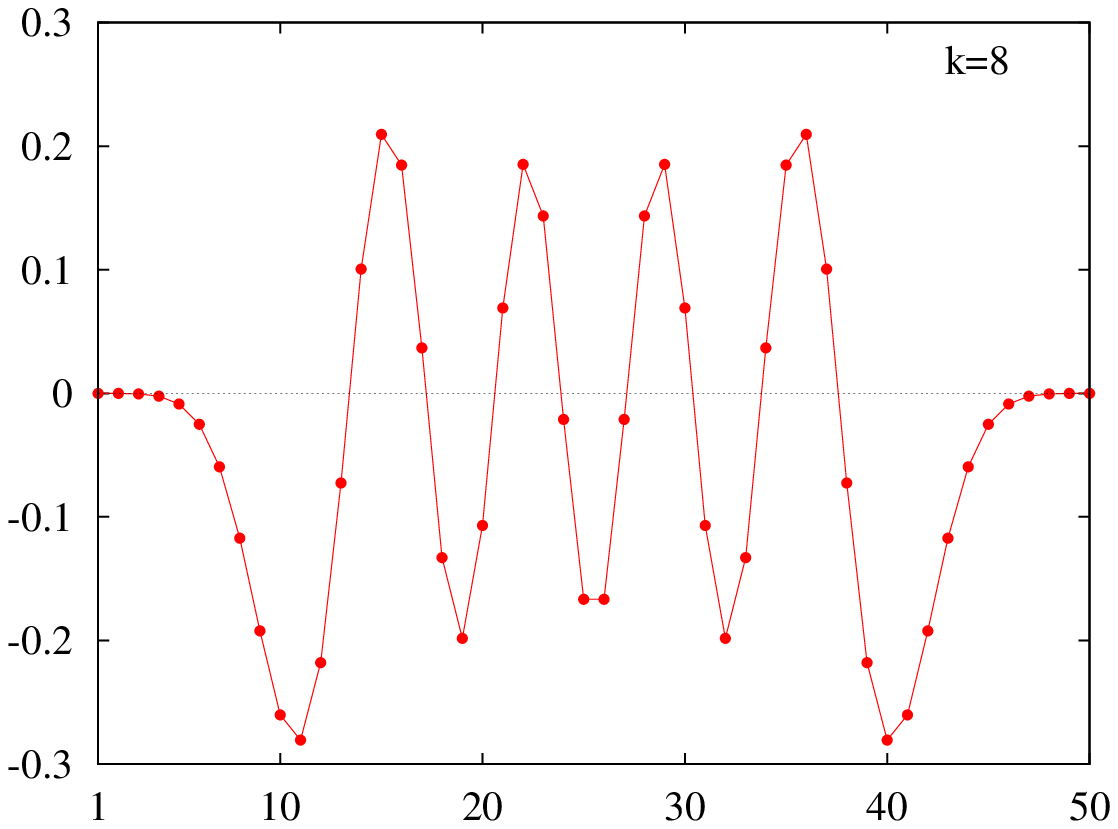}
\includegraphics[width=0.3\textwidth]{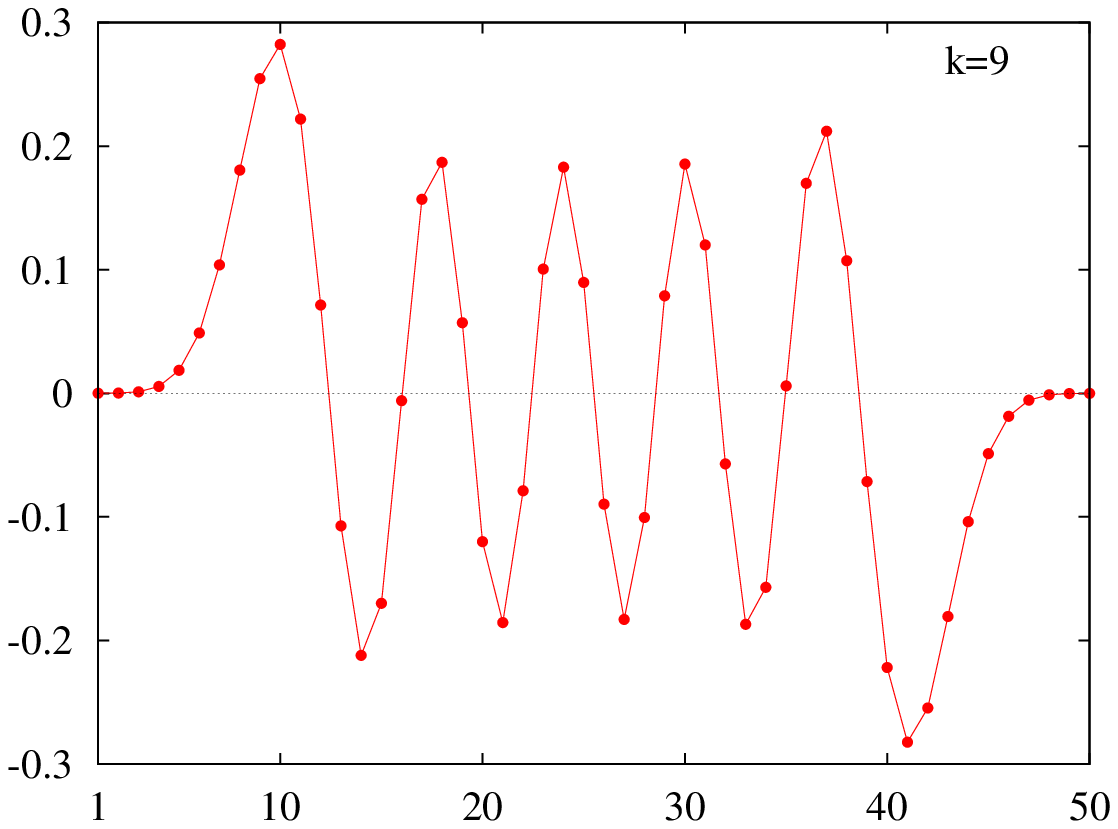}
\includegraphics[width=0.3\textwidth]{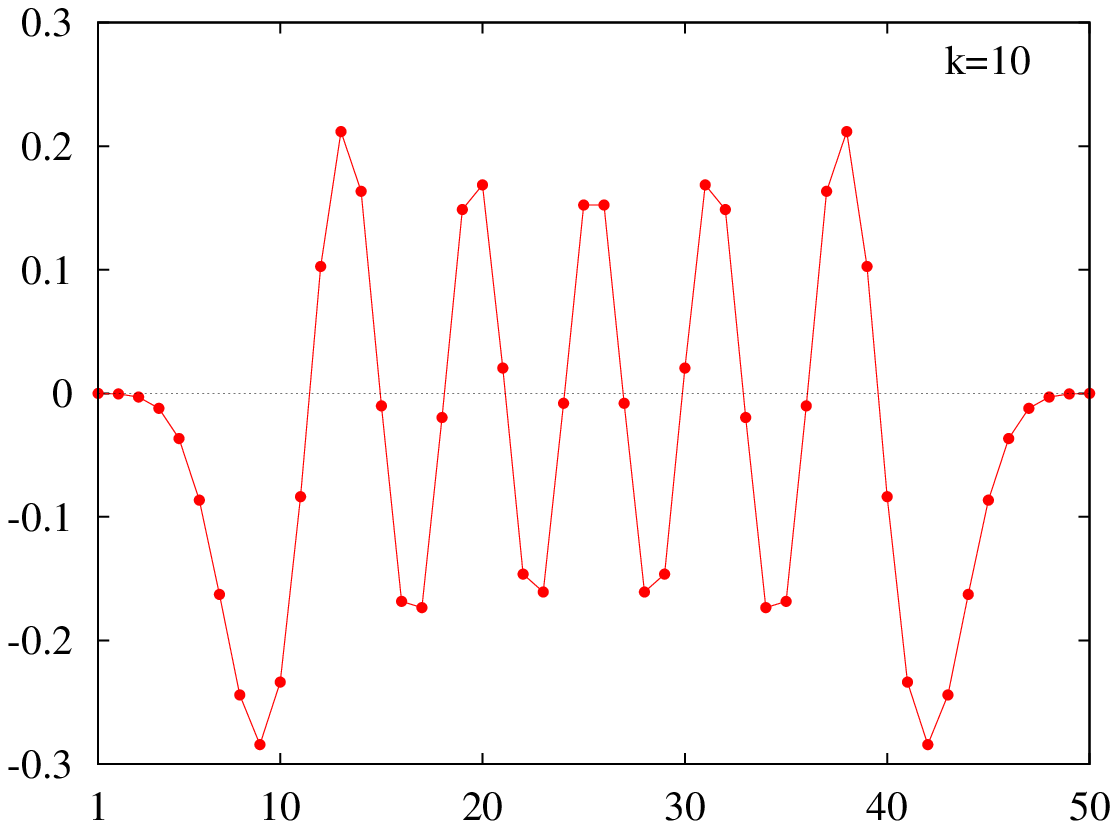}
\includegraphics[width=0.3\textwidth]{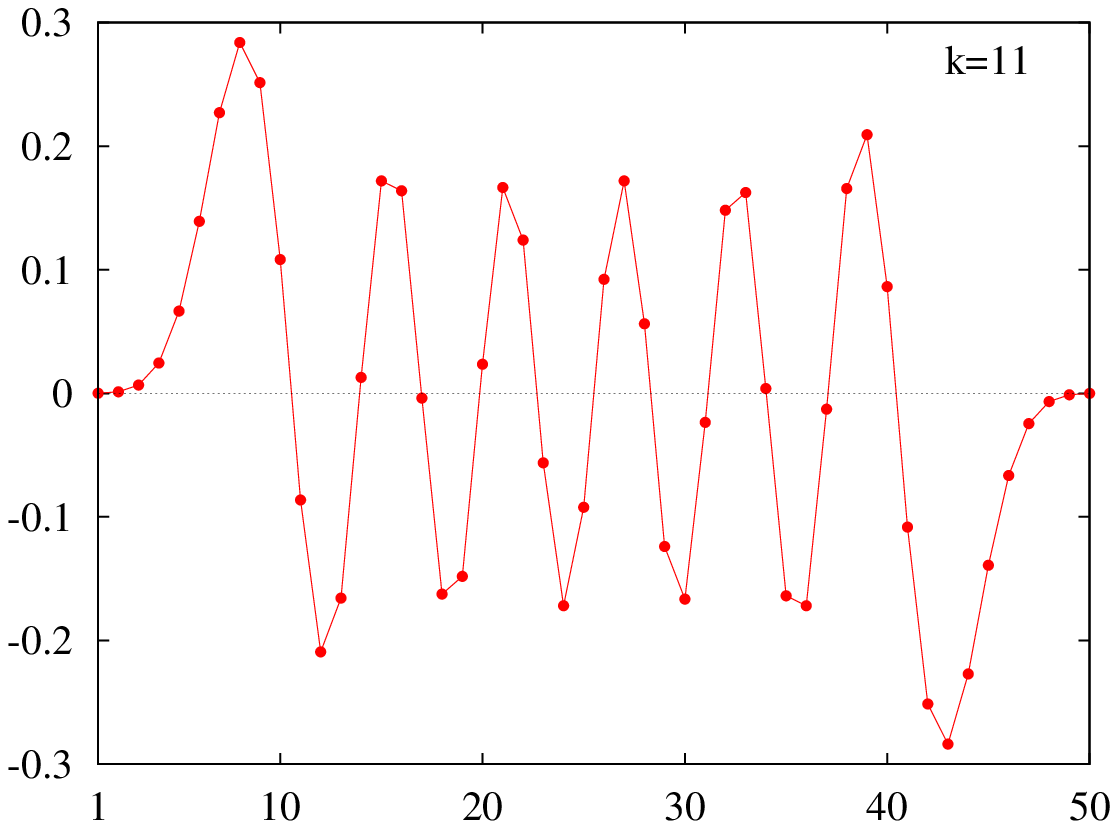}
\includegraphics[width=0.3\textwidth]{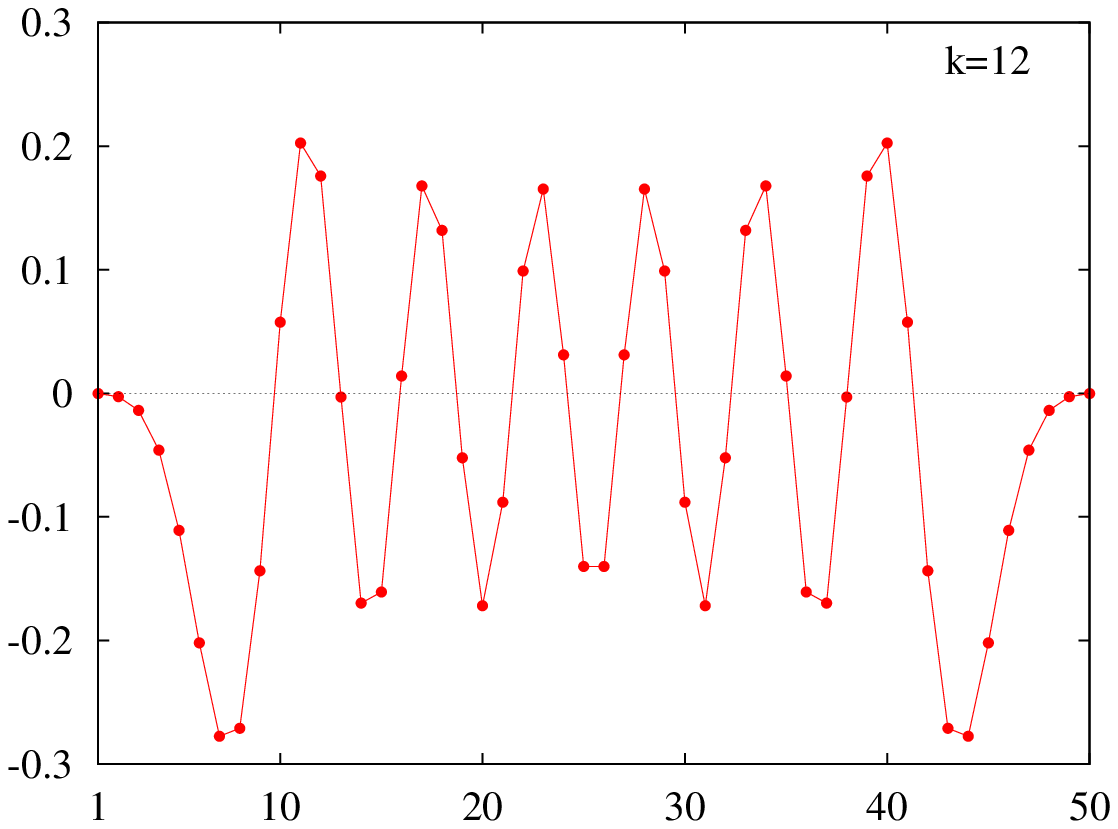}
\includegraphics[width=0.3\textwidth]{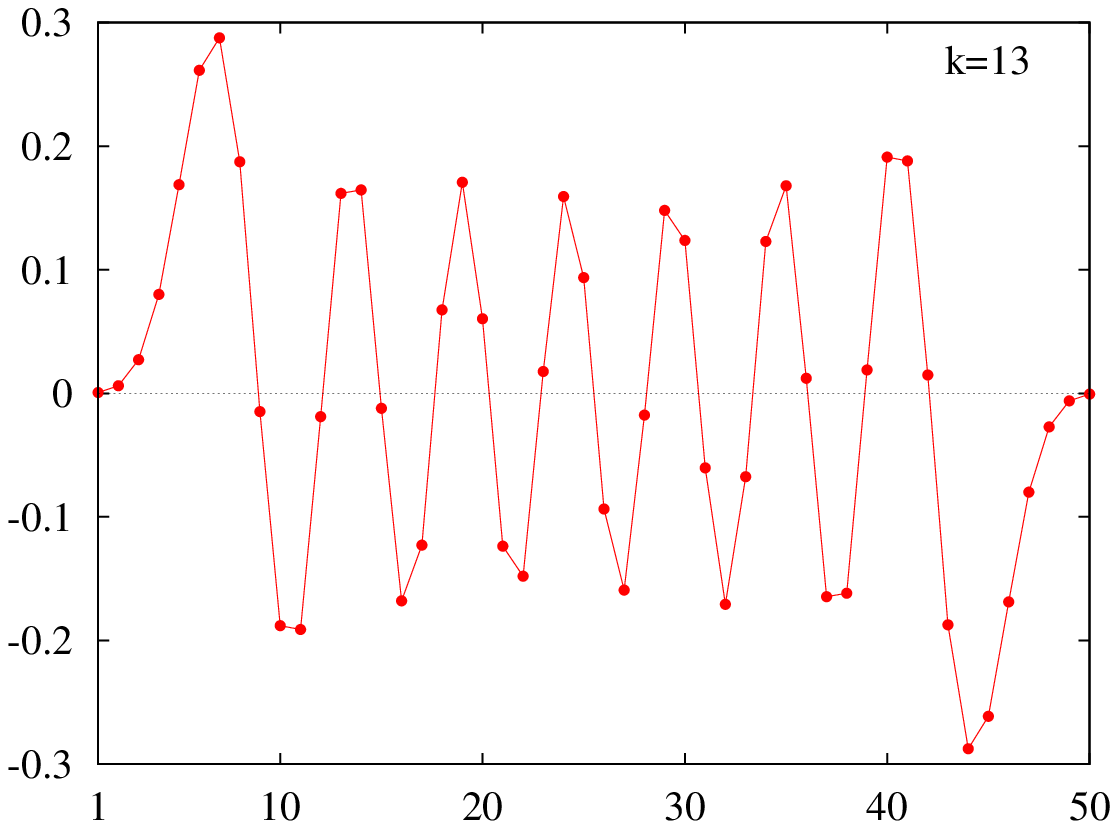}
\includegraphics[width=0.3\textwidth]{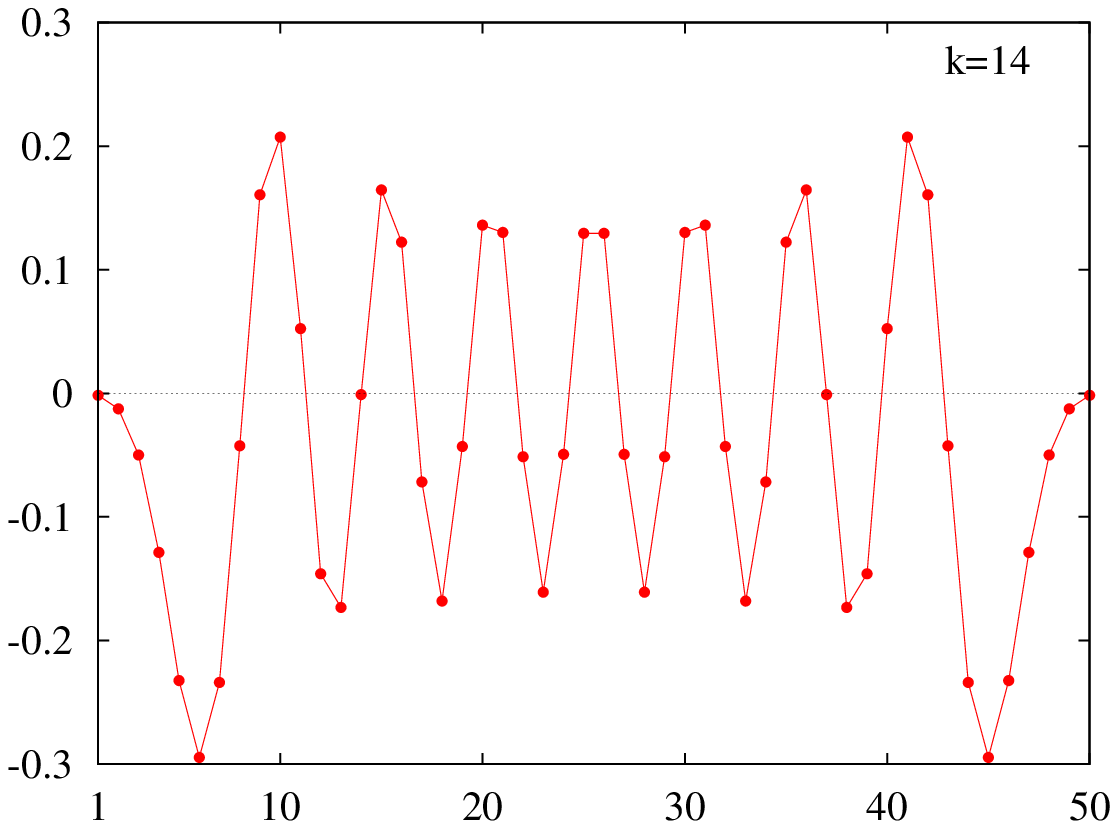}
\caption{The first 15 eigenfunctions $\varphi_k$ for a subsystem of $L=50$ sites 
in a half-filled infinite chain.}
\label{fig:lattice}
\end{figure}

This matrix was later rediscovered in \cite{Peschel04} on the basis that the Hamiltonian
in (\ref{rhogen}) when calculated numerically shows a very similar structure.
Note that $T$ is only determined up to a multiplicative factor and an additive constant.
The matrix elements given in \cite{Peschel04} differ from (\ref{tridiagonal2}) by the factor 
$2/L^2$ and a constant in $d_n$. Using $T$, one can obtain the eigenfunctions  $\varphi_k$
numerically for much larger systems than via the correlation matrix. They resemble the 
spheroidal functions of the continuum and were termed \emph{discrete prolate spheroidal
sequences} by Slepian. In Fig. \ref{fig:lattice} we show some of them for a system of $L=50$ 
sites and $q_F=\pi/2$. For $q_F=0,\pi$ they are given by Hahn polynomials and the eigenvalues of
$T$ are very simple \cite{Albanese04}.
In the limit $q_F \rightarrow 0$ and $L\rightarrow \infty$ keeping $q_FL$ constant, the 
problem becomes continuous in $y=n/L$ and one comes back to the usual spheroidal functions. 
Expanding the quantities in $T\varphi=\theta' \varphi$, one can also derive
the differential operator $D$ directly.

%
%

\begin{figure}[htb]
\center
\includegraphics[width=0.7\textwidth]{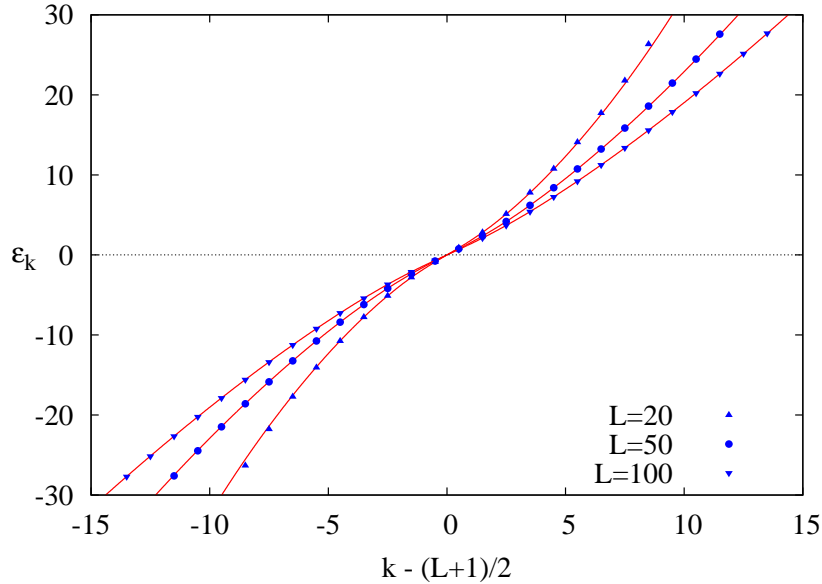}
\caption{Dispersion relation of the $\varepsilon_k$ for a half-filled infinite lattice and 
three different sizes of the subsystem. Symbols: numerical data, lines: analytical result for
$k(\varepsilon)$.}
\label{fig:dispersion}
\end{figure}

  The eigenvalues $\varepsilon_k$ follow directly from the formulae in section 4 with 
$N \rightarrow L$ and $2\pi W=q_F$. Then (\ref{epsilon_fin_as3}) becomes
\begin{equation}
\varepsilon_k = \frac{\pi^2(k- 1/2 - \bar N)}{\ln(2L \sin{\,q_F})-\psi(1/2)} 
\label{epsilon_latt_as}
\end{equation}
For a half-filled system ($q_F=\pi/2$) this gives the asymptotic result found in \cite{Peschel04},
whereas for $q_F \rightarrow 0$ and $L\rightarrow \infty$ it reduces to (\ref{epsilon_as3}).
The density of states $n(\varepsilon)$ which decreases with $\varepsilon$ is clearly seen in
numerical calculations of the $\varepsilon_k$ \cite{review09}. This is illustrated in 
Fig. \ref{fig:dispersion},
where the eigenvalues obtained directly from (\ref{corr_latt2}) are shown together with 
the theoretical formula for $k=k(\varepsilon)$, viewing $k$ as a continuous variable. The 
curves are not exactly linear, but show an upward (downward) bend on the positive (negative) 
side. One sees that for $L=20$
there are small differences between both results, but for the larger $L$ there is an
excellent agreement. One also sees that even there the linear region is still quite small. 

%
%

\section{Further aspects}

So far, we have considered only systems without boundaries. However, the results
cover also the case of a semi-infinite geometry when the subsystem is located
next to the boundary. The $\Phi_q$ are then sine functions and, in the continuum case,
lead to the correlation matrix
\begin{equation}
C_s(x,x') = C(x-x') - C(x+x') , \quad \quad x,x' \ge 0
\label{corr_semi_cont1}
\end{equation}
where $C(x-x')$ is the result (\ref{corr_cont1}). But if the subsystem is the interval
$0 \le x \le \ell/2$, the eigenvalue problem
\begin{equation}
\int_{0}^{\ell/2} dx' \, C_s(x-x') \varphi_k(x') = \zeta_k \,\varphi_k(x) 
\label{corr_semi_cont2}
\end{equation}
is solved by the \emph{odd} eigenfunctions of (\ref{corr_cont2}). Thus one can take over
the previous results, confining oneself to the odd $k$-values. This means in particular that
the spacing of the low-lying $\varepsilon_k$ doubles and the density of states halves.
As a result, the entanglement entropy is also halved.

The argument also holds for the lattice case, where $C_s$ has the form (\ref{corr_semi_cont1})
with $x,x' \rightarrow n,n'$ and $n,n' \ge 1$. However, the solutions of the analogue of 
(\ref{corr_semi_cont2}) then vanish at the point $n=0$. Therefore, one has to compare with an 
infinite lattice problem, where this point is the center of the subsystem. This means, that 
it has to consist of $2L+1$ sites (an odd number), if in the semi-infinite case the subsystem 
has $L$ sites. The connection between infinite and semi-infinite system is also seen at the
level of the commuting matrices. There exists a tridiagonal matrix $T_s$ with the property
$[C_s,T_s]=0$ and it is just the right half of the matrix $T$ \cite{Peschel04}.

Returning to closed systems, one should mention that a finite ring of $M$ lattice sites filled with
$N=2m+1$ particles leads to the correlation matrix (\ref{corr_cont_fin1}) with $x \rightarrow n$
and $L \rightarrow M$. For a segment of $L$ sites, one can then also find a commuting tridiagonal
matrix \cite{Grünbaum81} which is a generalization of $T$ and leads to another discrete version
of the spheroidal functions. The parabolic form of the $t_n$ and $d_n$ is then replaced by 
trigonometric expressions. For example $t_n \sim \sin(\pi n/M)\sin(\pi(L-n)/M)$, and for 
$M \rightarrow \infty$ one reobtains (\ref{tridiagonal2}).

Another generalization concerns higher dimensions. The case of an infinite continuum with spherical
Fermi surface and a spherical subsystem has also been studied in detail \cite{Slepian64}.
Due to the rotational symmetry, the eigenfunctions are products of an angular and a radial factor.
For the latter, one finds an integral equation with a Bessel function as kernel. Again a 
commuting differential operator exists and is closely related to $D$. For $d=2$ and with 
$y \rightarrow r$
\begin{equation}
D_r = D +\frac{m^2-1/4}{r^2}\,, \quad \quad r>0
\label{diff_op_2d}
\end{equation}
where $m = 0,\pm 1, \pm 2...$ is the azimuthal quantum number. Physically, the additional term can be
viewed as a centrifugal potential. As a result, the eigenfunctions have to vanish at the origin.
They were called \emph{generalized spheroidal functions} and some are shown in \cite{Slepian64}.
For the eigenvalues $\varepsilon_{km}$ for fixed $m$ one finds very similar results as in one 
dimension.

\section{Conclusion}

We have considered the ground state of free fermions and discussed an approach which
determines the reduced density matrix directly via its single-particle eigenfunctions. 
It circumvents the original matrices or integral kernels and works instead with differential 
operators. In the continuum case, this leads to spheroidal functions which are well known 
in mathematics and appear in various other areas. Physically, they have the character of
either bulk or boundary-like states, and it is amusing that, in the first instance, they 
are close to harmonic oscillator functions. In hindsight, however, one could have guessed 
that from the appearance of their lattice counterparts. In the lattice case, the asymptotic 
results for the single-particle eigenvalues bridge a gap between the numerical calculations 
and the Fisher-Hartwig determinantal formulae which only give a density of states. 
We have demonstrated in Fig. \ref{fig:dispersion} how well the results match.

On the technical level, the approach is rather tedious and inferior to Fisher-Hartwig 
calculations. However, the concept of a commuting operator is interesting and reminiscent
of transfer matrix problems in integrable classical lattice models.  It is also interesting 
that this feature carries over to higher dimensions for spherical subsystems. As to the 
one-dimensional problem, one has now an almost complete overview over the properties of the 
entanglement Hamiltonian $\mathcal{H}$, but an explicit form is still lacking.

\vspace{1cm}

\begin{acknowledgements}
We thank Pasquale Calabrese and Vladislav Popkov for interesting discussions. 
V.E. acknowledges financial support by the ERC grant QUERG.
\end{acknowledgements}

\section*{References}

\providecommand{\newblock}{}

\end{document}